\pdfoutput=1 
\documentclass[oneside,reqno,english]{amsart}
\usepackage[T1]{fontenc}
\usepackage[latin9]{inputenc}
\usepackage{geometry}
\usepackage{textcomp}
\usepackage{amsthm}
\usepackage{amssymb}

\makeatletter

\numberwithin{equation}{section}
\numberwithin{figure}{section}

\usepackage{etoolbox}
\usepackage[hidelinks]{hyperref}
\makeatletter
\patchcmd{\@maketitle}
  {\ifx\@empty\@dedicatory}
  {\ifx\@empty\@date \else {\vskip3ex \centering\footnotesize\@date\par\vskip1ex}\fi
   \ifx\@empty\@dedicatory}
  {}{}
\patchcmd{\@adminfootnotes}
  {\ifx\@empty\@date\else \@footnotetext{\@setdate}\fi}
  {}{}{}
\makeatother

\makeatother

\usepackage{babel}
\begin{document}
\title{A dynamic programming interpretation of Quantum Mechanics}
\author{Adam Brownstein$^{*}$}
\thanks{$^{*}$Melbourne, Australia. ORCID: https://orcid.org/0009-0001-7814-4384\\
}
\date{January 8, 2024}
\begin{abstract}
\noindent We introduce a transformation of the quantum phase $S'=S+\frac{\hbar}{2}\log\rho$,
which converts the deterministic equations of quantum mechanics into
the Lagrangian reference frame of stochastic particles. We show that
the quantum potential can be removed from the transformed quantum
Hamilton-Jacobi equations if they are solved as stochastic Hamilton-Jacobi-Bellman
equations. The system of equations provide a local description of
quantum mechanics, which is enabled by the inherently retrocausal
nature of stochastic Hamilton-Jacobi-Bellman equations. We also investigate
the stochastic transformation of the classical system, where is it
shown that quantum mechanics with the quantum potential reduced by
a factor of $\frac{1}{2}$ has a classical representation, which may
have interesting implications. Finally, we discuss the notion of a
subsystem correspondence principle, which constrains the ontology
of the total quantum system. 
\end{abstract}

\maketitle

\section{Introduction}

\noindent Dynamic programming is a computational method which uses
memoization and backwards induction to solve computational problems
with branching state space in linear time. The Hamilton-Jacobi-Bellman
equations provide a link between dynamic programming and the Hamilton-Jacobi
formalism of mechanics. The de Broglie-Bohm interpretation is based
on quantum Hamilton-Jacobi equations, and therefore can be understood
as a type of dynamic program. What is surprising is that a dynamic
programming interpretation of quantum mechanics is quite natural,
and may shed light on the ontology of quantum entanglement.

Conceptually, dynamic programming has remarkable similarities to quantum
mechanics. In dynamic programming problems, a value function is propagated
backward from the final rows of a computational grid toward the initial
row of the grid. After propagation of the value function, the computational
solution is found in the forward direction of the grid using local
information contained in the value function. The analogy to quantum
mechanics is that the dynamic programming value function is similar
to the quantum phase. In quantum mechanics, the particle phase stores
information of all possible particle paths, whereas the dynamic programming
value function stores information of all possible combinations of
the computational problem. Furthermore, the solution to the dynamic
program is analogous to a de Broglie-Bohm particle. In particular,
the computational solution is influenced by apparently non-local effects
due to the value function having been precomputed in the reverse direction
of the grid.

In light of these similarities, a dynamic programming interpretation
of quantum mechanics could provide an explanation of the quantum phase
and entanglement correlations. However, two key limitations are that
firstly, the interpretation would be retrocausal, and secondly, for
many-body systems the quantum phase is a configuration space entity
not a three-dimensional space entity. This paper attempts to solve
the second problem by describing propagation of the quantum phase
without the non-local quantum potential term, so that it can be understood
in three-dimensional space. 

Regarding the first problem (of retrocausality), Bell's theorem \cite{key-6}
has shown that no local-realist interpretations of quantum entanglement
are possible, and any interpretation requires a novel philosophical
loophole. Retrocausality is one of the few viable loopholes available
to evade Bell's theorem. We accept retrocausality as a possible solution,
and highlight the analogy between retrocausal de Broglie-Bohm particles
and the method of backward-induction used to solve dynamic programming
problems.

To provide some perspective on the second problem (of configuration
space), the most popular interpretations of quantum mechanics are
all impacted by the configuration space description of the wavefunction
in some way. For example, the Copenhagen interpretation describes
the wavefunction in Hilbert space, which in the basis of particle
coordinates is a configuration space. For the de Broglie-Bohm interpretation,
the phase is a non-separable function in configuration space and the
many-body guidance equations are non-local. In the many-worlds interpretation,
the wavefunction is embedded in three-dimensional space only by describing
the state space as parallel outcomes which simultaneously exist. Finally,
the path integral interpretation takes the sum over configuration
space paths. Therefore understanding the configuration space nature
of entanglement is evidently central to developing an interpretation
of quantum mechanics, and impacts all of these previous interpretations. 

This paper attempts to describe quantum mechanics in a three-dimensional
space ontology by transforming the quantum Hamilton-Jacobi equations
into the Lagrangian picture of stochastic particles. This transformation
enables the quantum Hamilton-Jacobi equations to be solved as stochastic
Hamilton-Jacobi-Bellman equations i.e. as a dynamic program. The quantum
potential term vanishes in the stochastic Hamilton-Jacobi-Bellman
picture due to the divergence theorem. Without the quantum potential
term, the equations of phase propagation can be interpreted locally
in three-dimensional space. 

As an outline of this paper, in section \ref{sec:de-Broglie-Bohm-interpretation},
the de Broglie-Bohm interpretation and an early stochastic interpretation
developed by Bohm, Vigier \& Hiley \cite{key-1,key-2} are presented.
In section \ref{sec:A-conceptual-framework}, we show how generic
Hamilton-Jacobi equations can be transformed into stochastic Hamilton-Jacobi-Bellman
equations where the non-local divergence terms vanish. In section
\ref{sec:Stochastic-transform-of}, we apply this transformation to
the quantum Hamilton-Jacobi equation, which enables the quantum potential
term to be eliminated. In section \ref{sec:Local-interpretation-with},
we discuss particle and phase propagation in three-dimensional space.
In section \ref{sec:From particles to policies}, we map several concepts
from dynamic programming to quantum mechanics in the de Broglie-Bohm
interpretation. In section \ref{sec:Alternative-transformation-of},
we provide an alternative transformation of the quantum Hamilton-Jacobi
equation based on retrocausal particle diffusion. In section \ref{sec:Stochastic-transformations-of},
we apply the transformations developed in sections \ref{sec:Stochastic-transform-of}
and \ref{sec:Alternative-transformation-of} to the classical Hamilton-Jacobi
equation, and demonstrate a connection to Nelson's stochastic mechanics
\cite{key-1-1}. In section \ref{sec:Correspondence-principle-of}
we introduce the notion of a subsystem correspondence principle, which
draws attention to issues in the ontological interpretation of quantum
subsystems.

\section{\label{sec:de-Broglie-Bohm-interpretation}De Broglie-Bohm interpretation}

\noindent This section describes the de Broglie-Bohm interpretation,
starting from the single-particle case and progressing to the many-particle
case. A stochastic version of the de Broglie-Bohm interpretation is
also presented. The many-particle case and stochastic interpretation
provide useful background for the subsequent results of the paper. 

\subsection{Single particle case}

The de Broglie-Bohm interpretation has two main equations, which are
derived from the real and imaginary components of the Schr{\"o}dinger
equation. The first equation is the quantum Hamilton-Jacobi equation,
which describes the time evolution of the quantum phase:
\begin{equation}
\frac{\partial S({\bf x},t)}{\partial t}+\frac{1}{2m}\nabla S({\bf x},t)\cdot\nabla S({\bf x},t)+V({\bf x},t)+Q({\bf x},t)=0,\label{eq:1}
\end{equation}
where the quantum potential term $Q({\bf x},t)$ is equal to: 
\begin{equation}
Q({\bf x},t)=-\frac{\hbar^{2}}{2m}\frac{\nabla^{2}\sqrt{\rho({\bf x},t)}}{\sqrt{\rho({\bf x},t)}}.
\end{equation}
The second equation is the continuity equation, which describes the
time evolution of the quantum probability density $\rho({\bf x},t)$:
\begin{equation}
\frac{\partial\rho({\bf x},t)}{\partial t}+\nabla\cdot\left(\rho({\bf x},t)\frac{\nabla S({\bf x},t)}{m}\right)=0.\label{eq:2}
\end{equation}
The logic of the de Broglie-Bohm interpretation is that particles
actually exist, and have an equation of motion given by the guidance
equation {[}Eq. \ref{eq:3}{]}: 
\begin{equation}
\frac{d{\bf X}(t)}{dt}=\left.\frac{\nabla S({\bf x},t)}{m}\right|_{{\bf x}={\bf X}(t)},\label{eq:3}
\end{equation}
where the conditioning symbol $|_{{\bf x}={\bf X}(t)}$ indicates
that the function is evaluated at the actual location of the particle
${\bf X}(t)$. 

If a statistical ensemble of particles follow this guidance equation,
the particle distribution is guaranteed to reproduce the quantum probability
density due to the equivariance property. Equivariance occurs because
the chosen guidance equation ensures the particle distribution follows
a continuity equation identical in form to the continuity equation
for the quantum probability density. The important point here is that
there are two separate continuity equations, one for the particle
distribution and another for the quantum mechanical distribution.
Because a continuity equation is a first-order differential equation
in time, its solutions are unique. Therefore the particle distribution
and quantum probability density must remain equal if they are equal
at the initial time. 

\subsection{Many particle case}

The generalisation of the de Broglie-Bohm interpretation to the case
of multiple particles is very natural. The coordinates of the single
particle get upgraded to the coordinates of a collection of particles,
which is alternatively described as a single point in configuration
space \cite{key-5}. The derivative operator is replaced by a tuple
of derivative operators. We will represent these multipartite generalisations
in a tuple-of-tuples notation as follows:
\begin{align}
\nabla\equiv & (\nabla_{1},\nabla_{2},...,\nabla_{n}) & {\bf x}\equiv & (x_{1},x_{2},...,x_{n}) & {\bf X}\equiv & (X_{1},X_{2},...,X_{n}),
\end{align}
where $X_{i}$ are particle coordinates and $x_{i}$ are grid coordinates
in three-dimensional space. For illustration, in the bipartite case,
the quantum Hamilton-Jacobi equation becomes:
\begin{equation}
\frac{\partial S({\bf x},t)}{\partial t}+\frac{1}{2m}{\bf \nabla}S({\bf x},t)\cdot{\bf \nabla}S({\bf x},t)+V({\bf x},t)+Q({\bf x},t)=0,\label{eq:6-1}
\end{equation}
\noindent which expanded in terms of individual particles coordinates
is equivalent to:
\begin{align}
\frac{\partial S(x_{1},x_{2},t)}{\partial t}+\frac{1}{2m}\nabla_{1}S(x_{1},x_{2},t)\cdot\nabla_{1}S(x_{1},x_{2},t)+\frac{1}{2m}\nabla_{2}S(x_{1},x_{2},t)\cdot\nabla_{2}S(x_{1},x_{2},t)\nonumber \\
+V(x_{1},x_{2},t)+Q(x_{1},x_{2},t) & =0.
\end{align}
\noindent Similarly, the quantum potential in the bipartite case
is equal to:
\begin{align}
Q({\bf x},t)= & -\frac{\hbar^{2}}{2m}\frac{\nabla^{2}\sqrt{\rho({\bf x},t)}}{\sqrt{\rho({\bf x},t)}}\\
= & -\frac{\hbar^{2}}{2m}\frac{\nabla_{1}^{2}\sqrt{\rho(x_{1},x_{2},t)}}{\sqrt{\rho(x_{1},x_{2},t)}}-\frac{\hbar^{2}}{2m}\frac{\nabla_{2}^{2}\sqrt{\rho(x_{1},x_{2},t)}}{\sqrt{\rho(x_{1},x_{2},t)}}.
\end{align}
 \noindent The many-body guidance equation in the general multipartite
case is:
\begin{equation}
\frac{dX_{i}(t)}{dt}=\left.\frac{\nabla_{i}S({\bf x},t)}{m}\right|_{{\bf x}={\bf X}(t)}.\label{eq:}
\end{equation}
As indicated in this formula, the guidance equation for particle $i$
is evaluated at the location of the particle configuration ${\bf X}(t),$
which is represented by the conditioning subscript $|_{{\bf x}={\bf X}(t)}$.
This evaluation procedure is non-local, and is the source of entanglement
correlations in the de Broglie-Bohm interpretation.

\subsection{\label{subsec:Stochastic-de-Broglie-Bohm}Stochastic de Broglie-Bohm
interpretation}

As recognized by Bohm, Vigier \& Hiley \cite{key-1,key-2}, de Broglie-Bohm
particles can be stochastic and still reproduce the continuity equation
of quantum mechanics. Consider adding a term $\frac{k}{2m}\nabla^{2}\rho({\bf x},t)$
to both sides of the continuity equation. The result is the following
equation:
\begin{equation}
\frac{\partial\rho({\bf x},t)}{\partial t}+{\bf \nabla}\cdot\left[\rho({\bf x},t)\left(\frac{1}{m}{\bf {\bf \nabla}}S({\bf x},t)+\frac{k}{2m}{\bf {\bf \nabla}}\log\rho({\bf x},t)\right)\right]=\frac{k}{2m}{\bf \nabla}^{2}\rho({\bf x},t),\label{eq:4}
\end{equation}
where $\nabla\log\rho({\bf x},t)=\frac{\nabla\rho({\bf x},t)}{\rho({\bf x},t)}$
has been used to simplify the left-hand side expression. This equation
can be further simplified by transforming the phase as $S'({\bf x},t)=S({\bf x},t)+\frac{k}{2}\log\rho({\bf x},t)$:
\begin{equation}
\frac{\partial\rho({\bf x},t)}{\partial t}+{\bf \nabla}\cdot\left(\rho({\bf x},t)\frac{{\bf \nabla}S'({\bf x},t)}{m}\right)=\frac{k}{2m}{\bf \nabla}^{2}\rho({\bf x},t).\label{eq:12}
\end{equation}
The transformed continuity equation is equivalent to a Fokker-Planck
equation for particles undergoing a Brownian motion, with drift velocity
and stochastic velocity terms given by the following guidance equation:
\begin{equation}
d{\bf X}(t)=\frac{\nabla S'({\bf x},t)}{m}dt+\sqrt{\frac{k}{m}}d{\bf W}_{t},\label{eq:13-1-1}
\end{equation}

\medskip{}
\noindent where $d{\bf W}_{t}$ is a Wiener process with $E\left[dW_{t}^{i}dW_{t}^{j}\right]=\delta^{ij}dt$.
An ensemble of de Broglie-Bohm particle configurations which have
this guidance equation will reproduce the quantum predictions due
to the equivariance property. 

\section{\label{sec:A-conceptual-framework}Non-local Hamilton-Jacobi equations
in the Lagrangian picture}

\noindent In this section, we demonstrate that a particular form
of non-local Hamilton-Jacobi equation in the Eulerian picture can
be described locally when transformed to the Lagrangian picture of
stochastic particles. This is the conceptual foundation for section
\ref{sec:Stochastic-transform-of}, where it is shown that the non-local
terms in the quantum Hamilton-Jacobi equation conform to this type
of reducible non-locality. 

\subsection{A non-local Hamilton-Jacobi equation in the Lagrangian picture}

Suppose there is a particle configuration ${\bf X}(t)$ which has
a stochastic guidance equation of the form described in section \ref{subsec:Stochastic-de-Broglie-Bohm}:
\begin{equation}
d{\bf X}(t)=\frac{{\bf \nabla}S'({\bf x},t)}{m}dt+\sqrt{\frac{k}{m}}d{\bf W}_{t},\label{eq:6}
\end{equation}
Assume that the phase $S'({\bf x},t)$ satisfies a modified Hamilton-Jacobi
equation containing a non-local term ${\bf \nabla}^{2}f({\bf x},t)$
and an additional $\frac{k}{2m}{\bf \nabla}^{2}S'({\bf x},t)$ term:
\begin{equation}
\frac{\partial S'({\bf x},t)}{\partial t}+\frac{1}{2m}{\bf \nabla}S'({\bf x},t)\cdot{\bf \nabla}S'({\bf x},t)+\frac{k}{2m}{\bf \nabla}^{2}S'({\bf x},t)+V({\bf x},t)+{\bf \nabla}^{2}f({\bf x},t)=0.\label{eq:7}
\end{equation}

\medskip{}
\noindent This Hamilton-Jacobi equation can be rearranged into the
following form: 
\begin{equation}
\left[\frac{\partial}{\partial t}+\frac{1}{m}{\bf \nabla}S'({\bf x},t)\cdot{\bf \nabla}+\frac{k}{2m}{\bf \nabla}^{2}\right]S'({\bf x},t)=\frac{1}{2m}{\bf \nabla}S'({\bf x},t)\cdot{\bf \nabla}S'({\bf x},t)-V({\bf x},t)-{\bf \nabla}^{2}f({\bf x},t).\label{eq:7-1}
\end{equation}

\medskip{}
\noindent Using the definition of the retrocausal stochastic Lagrangian
derivative (Appendix \ref{subsec:Retrocausal-Lagrangian-derivativ}),
{[}Eq. \ref{eq:7-1}{]} becomes:
\begin{equation}
\frac{D_{*}S'({\bf x},t)}{Dt}=\frac{1}{2m}{\bf \nabla}S'({\bf x},t)\cdot{\bf \nabla}S'({\bf x},t)-V({\bf x},t)-{\bf \nabla}^{2}f({\bf x},t).\label{eq:8-2}
\end{equation}

\subsection{\label{subsec:Elimination-of-the}Elimination of the non-local term}

The $\nabla^{2}f(x)$ term of {[}Eq. \ref{eq:8-2}{]} can be eliminated
by using the divergence theorem. To see this, firstly rearrange the
definition of the stochastic Lagrangian derivative in terms of $S'({\bf x},t)$: 

\begin{align}
E_{\delta{\bf x}}\left[\frac{S'({\bf x}+\delta{\bf x},t+\delta t)-S'({\bf x},t)}{\delta t}\right]=\frac{D_{*}S'({\bf x},t)}{Dt}\\
\Rightarrow S'({\bf x},t)=E_{\delta{\bf x}}\left[S'({\bf x}+\delta{\bf x},t+\delta t)\right]-\frac{D_{*}S'({\bf x},t)}{Dt}\delta t.\label{eq:38-1}
\end{align}
The phase one time-step ahead can be described analogously to {[}Eq.
\ref{eq:38-1}{]} as:
\begin{equation}
S'({\bf x}+\delta{\bf x},t+\delta t)=E_{\delta{\bf y}}\left[S'({\bf x}+\delta{\bf x}+\delta{\bf y},t+2\delta t)\right]-\frac{D_{*}S'({\bf x}+\delta{\bf x},t+\delta t)}{Dt}\delta t.\label{eq:3.7}
\end{equation}
Consequently, we can expand {[}Eq. \ref{eq:38-1}{]} recursively by
substituting in {[}Eq. \ref{eq:3.7}{]}. For instance, expanding twice
gives:
\begin{align}
S'({\bf x},t)= & E_{\delta{\bf x}}\left[E_{\delta{\bf y}}\left[S'({\bf x}+\delta{\bf x}+\delta{\bf y},t+2\delta t)\right]\right]-E_{\delta{\bf x}}\left[\frac{D_{*}S'({\bf x}+\delta{\bf x},t+\delta t)}{Dt}\right]\delta t-\frac{D_{*}S'({\bf x},t)}{Dt}\delta t.\label{eq:11-4}
\end{align}
At this point, the equation for the Lagrangian derivative of the phase
{[}Eq. \ref{eq:8-2}{]} can be substituted into {[}Eq. \ref{eq:11-4}{]}.
For example, substitution into the second term gives:
\begin{align}
 & E_{\delta{\bf x}}\left[\frac{D_{*}S'({\bf x}+\delta{\bf x},t+\delta t)}{Dt}\right]\delta t\nonumber \\
 & =E_{\delta{\bf x}}\left[\frac{1}{2m}{\bf \nabla}S'({\bf x}+\delta{\bf x},t+\delta t)\cdot{\bf \nabla}S'({\bf x}+\delta{\bf x},t+\delta t)-V({\bf x}+\delta{\bf x},t+\delta t)-{\bf \nabla}^{2}f({\bf x}+\delta{\bf x},t+\delta t)\right]\delta t.
\end{align}
If the expectation value $E_{\delta{\bf x}}\left[\nabla^{2}f({\bf x}+\delta{\bf x},t+\delta t)\right]$
equals zero, then the non-local component is removed by the results
of section \ref{subsec:Vanishing-expectation-of}, and {[}Eq. \ref{eq:11-4}{]}
becomes:
\begin{align}
S'({\bf x},t)= & E_{\delta{\bf x}}\left[E_{\delta{\bf y}}\left[S'({\bf x}+\delta{\bf x}+\delta{\bf y},t+2\delta t)\right]\right]-E_{\delta{\bf x}}\left[\mathcal{L}_{c}({\bf x}+\delta{\bf x},t+\delta t)\right]\delta t-\frac{D_{*}S'({\bf x},t)}{Dt}\delta t,\label{eq:11-4-1}
\end{align}
where $\mathcal{L}_{c}$ is the classical Lagrangian:
\begin{equation}
\mathcal{L}_{c}({\bf x},t)=\frac{1}{2m}{\bf \nabla}S'({\bf x},t)\cdot{\bf \nabla}S'({\bf x},t)-V({\bf x},t).
\end{equation}
This equational form {[}Eq. \ref{eq:11-4-1}{]} remains true if we
perform subsequent expansions, provided the expectation values over
the non-local terms e.g.{\small{} $E_{\delta{\bf z}}\left[\nabla^{2}f({\bf x}+\delta{\bf x}+...+\delta{\bf z},t+\delta t+...+\delta t)\right]$}
all equal zero. For example, expanding three times gives:
\begin{align}
 & S'({\bf x},t)=E_{\delta{\bf x}}\left[E_{\delta{\bf y}}\left[E_{\delta{\bf z}}\left[S'({\bf x}+\delta{\bf x}+\delta{\bf y}+\delta{\bf z},t+3\delta t)\right]\right]\right]-E_{\delta{\bf x}}\left[E_{\delta{\bf y}}\left[\mathcal{L}_{c}({\bf x}+\delta{\bf x}+\delta{\bf y},t+2\delta t)\right]\right]\delta t\nonumber \\
 & -E_{\delta{\bf x}}\left[\mathcal{L}_{c}({\bf x}+\delta{\bf x},t+\delta t)\right]\delta t-\frac{D_{*}S'({\bf x},t)}{Dt}\delta t.\label{eq:25}
\end{align}
While the first term of the expansion $\frac{DS'({\bf x},t)}{Dt}\delta t$
is potentially non-local, this initial term only has an infinitesimal
effect upon $S'({\bf x},t)$ and can be ignored. The remaining terms
are all local. Note that equations {[}Eq. \ref{eq:38-1}{]} and {[}Eq.
\ref{eq:25}{]} are stochastic Hamilton-Jacobi-Bellman equations,
thus providing a link to dynamic programming. 

\subsection{\label{subsec:Vanishing-expectation-of}Divergence theorem of the
non-local term}

In the previous section, we have made the assumption that: 
\begin{equation}
E_{\delta{\bf x}}\left[\nabla^{2}f({\bf x}+\delta{\bf x},t+\delta t)\right]\approx0.
\end{equation}
To show this ansatz is true, we firstly write the expectation as a
sum over particle configurations with coordinates $q_{a}=({\bf x}+\delta{\bf x},t+\delta t)$:
\begin{align}
E_{\delta{\bf x}}\left[\nabla^{2}f({\bf x}+\delta{\bf x},t+\delta t)\right]= & \sum_{a}\nabla^{2}f(q_{a})\label{eq:35}
\end{align}
The sum over particle configurations can be understood in terms of
probabilities for particles to transition from their initial coordinates
to their final coordinates. Therefore {[}Eq. \ref{eq:35}{]} can be
written in terms of transition probabilities of particles beginning
at the point $({\bf x},t)$ and ending at the collection of points
$({\bf x}+\delta{\bf x},t+\delta t)$ as follows: 
\begin{align}
E_{\delta{\bf x}}\left[\nabla^{2}f({\bf x}+\delta{\bf x},t+\delta t)\right]= & \sum_{a}\nabla^{2}f(q_{a})\\
= & \sum_{{\bf \delta{\bf x}}}P({\bf x}+\delta{\bf x},t+\delta t|{\bf x},t)P({\bf x},t)\nabla_{{\bf x}+\delta{\bf x}}^{2}f({\bf x}+\delta{\bf x},t+\delta t)\\
\approx & \int P({\bf y},t+\delta t|{\bf x},t)P({\bf x},t)\nabla_{{\bf y}}^{2}f({\bf y},t+\delta t)d{\bf y}\label{eq:30-1}\\
\approx & \left[\int P({\bf y},t+\delta t|{\bf x},t)\nabla_{{\bf y}}^{2}f({\bf y},t+\delta t)d{\bf y}\right]P({\bf x},t),
\end{align}
where ${\bf x}+\delta{\bf x}$ has been replaced with the variable
${\bf y}$ for simplicity of notation, and the sum $\sum_{{\bf \delta{\bf x}}}$
is approximated as an integral in the continuous limit. The derivative
operator $\nabla$ can now shifted onto the $P({\bf y},t+\delta t|{\bf x},t)$
term by using the inverse product rule and applying the divergence
theorem twice, which gives: 
\begin{align}
E_{\delta{\bf x}}\left[\nabla^{2}f({\bf x}+\delta{\bf x},t+\delta t)\right]\approx & \left[\int f({\bf y},t+\delta t)\nabla_{{\bf y}}^{2}P({\bf y},t+\delta t|{\bf x},t)d{\bf y}\right]P({\bf x},t)\label{eq:31-1}
\end{align}
The function $P({\bf y},t+\delta t|{\bf x},t)$ in {[}Eq. \ref{eq:31-1}{]}
describes the transition probability for particles to move from $({\bf x},t)$
to $({\bf y},t+\delta t)$. However, the particles can only move a
short distance over small time horizons, therefore $P({\bf y},t+\delta t|{\bf x},t)\approx0$
when ${\bf y}$ is materially different from ${\bf x}$. Consequently,
$\nabla_{{\bf y}}^{2}P({\bf y},t+\delta t|{\bf x},t)\approx0$ except
in a small region of values around ${\bf x}$. Because the function
$f({\bf y},t+\delta t)$ is approximately a constant in this small
region of values, we can take this term outside the integral as a
constant. This results in: 
\begin{align}
E_{\delta{\bf x}}\left[\nabla^{2}f({\bf x}+\delta{\bf x},t+\delta t)\right]\approx & \left[\int f({\bf y},t+\delta t)\nabla_{{\bf y}}^{2}P({\bf y},t+\delta t|{\bf x},t)d{\bf y}\right]P({\bf x},t)\\
\approx & \left[\int\nabla_{{\bf y}}^{2}P({\bf y},t+\delta t|{\bf x},t)d{\bf y}\right]f({\bf y},t+\delta t)P({\bf x},t)\label{eq:34}\\
\approx & 0,\label{eq:36}
\end{align}
where we have used the divergence theorem to equate {[}Eq. \ref{eq:34}{]}
to zero. 

\subsection{Retrocausality of the stochastic Hamilton-Jacobi-Bellman equations}

In light of Bell's theorem, it is peculiar that the non-local terms
can be removed from the stochastic Hamilton-Jacobi-Bellman equations.
This is made possible due to the recursive expansion of the phase
$S({\bf x},t)$ being conducted in the backward direction of time.
Stochastic Hamilton-Jacobi-Bellman equations are naturally solved
in the opposite temporal direction to the direction of the stochastic
particle frame, using backward induction to perform the physical computation. 

To understand the retrocausal nature of stochastic Hamilton-Jacobi-Bellman
equations further, imagine the particle distribution is stochastic
in the forward direction of time as per usual. This means that a collection
of particles will spread out from an initial point to a range of final
points. The stochastic Hamilton-Jacobi-Bellman equations indicate
the initial phase can be determined by tracing the collection of particles
backward along their trajectories, reversing the update of the phase
along each particle path, then taking the average of the result. But
the particle trajectories only meet at single spacetime coordinate
at the earlier time, so the calculation naturally takes place in the
reverse direction of time.

\section{\label{sec:Stochastic-transform-of}Stochastic transformation of
the quantum Hamilton-Jacobi equation}

\noindent In this section, we transform the quantum Hamilton-Jacobi
equations into the Lagrangian reference frame of stochastic particles.
This transformation enables the quantum Hamilton-Jacobi equations
to be written as stochastic Hamilton-Jacobi-Bellman equations. It
is shown that the non-local quantum potential term can be removed
from the stochastic Hamilton-Jacobi-Bellman picture using the method
developed in section \ref{sec:A-conceptual-framework}. 

\subsection{Transformed quantum Hamilton-Jacobi equations\label{subsec:Transformed-quantum-Hamilton-Jac} }

Firstly, the quantum Hamilton-Jacobi equation in standard form is:
\begin{equation}
\frac{\partial S}{\partial t}+\frac{1}{2m}\nabla S\cdot\nabla S+V+Q=0.\label{eq:1-1-3}
\end{equation}
Note that the quantum potential $Q$ can be written as {[}Eq. \ref{eq:113}{]}:
\begin{align}
Q= & \frac{\hbar^{2}}{8m}\nabla\log\rho\cdot\nabla\log\rho-\frac{\hbar^{2}}{4m}\frac{\nabla^{2}\rho}{\rho}.\label{eq:10}
\end{align}
Substitute this form of the quantum potential into {[}Eq. \ref{eq:1-1-3}{]}
to give:
\begin{align}
\frac{\partial S}{\partial t}+\frac{1}{2m}\nabla S\cdot\nabla S+V+\frac{\hbar^{2}}{8m}\nabla\log\rho\cdot\nabla\log\rho-\frac{\hbar^{2}}{4m}\frac{\nabla^{2}\rho}{\rho}=0.
\end{align}

\medskip{}
\noindent Add and subtract several terms:
\begin{align}
\frac{\partial S}{\partial t}+\frac{1}{2m}\nabla S\cdot\nabla S+V+\left[\frac{\hbar}{2m}\nabla^{2}S-\frac{\hbar}{2m}\nabla^{2}S\right]+\left[\frac{\hbar^{2}}{4m}\nabla^{2}\log\rho-\frac{\hbar^{2}}{4m}\nabla^{2}\log\rho\right]\nonumber \\
+\left[\frac{\hbar}{2m}\nabla\log\rho\cdot\nabla S-\frac{\hbar}{2m}\nabla\log\rho\cdot\nabla S\right]+\frac{\hbar^{2}}{8m}\nabla\log\rho\cdot\nabla\log\rho-\frac{\hbar^{2}}{4m}\frac{\nabla^{2}\rho}{\rho} & =0.\label{eq:4.4}
\end{align}

\medskip{}
\noindent Rearrange the resulting equation:
\begin{align}
\left[\frac{\partial S}{\partial t}-\frac{\hbar}{2m}\nabla^{2}S-\frac{\hbar}{2m}\nabla S\cdot\nabla\log\rho\right]+\left[\frac{1}{2m}\nabla S\cdot\nabla S+\frac{\hbar}{2m}\nabla\log\rho\cdot\nabla S+\frac{\hbar^{2}}{8m}\nabla\log\rho\cdot\nabla\log\rho\right]\nonumber \\
+\left[\frac{\hbar}{2m}\nabla^{2}S+\frac{\hbar^{2}}{4m}\nabla^{2}\log\rho\right]+\left[-\frac{\hbar^{2}}{4m}\nabla^{2}\log\rho-\frac{\hbar^{2}}{4m}\frac{\nabla^{2}\rho}{\rho}\right] & =0.\label{eq:31}
\end{align}

\medskip{}
\noindent Using the stochastic de Broglie-Bohm interpretation (section
\ref{subsec:Stochastic-de-Broglie-Bohm}) with diffusion constant
$k=\hbar$, the particle configuration can be assumed to have a guidance
equation of the form:
\begin{equation}
d{\bf X}(t)=\frac{1}{m}{\bf \nabla}S'({\bf x},t)dt+\sqrt{\frac{\hbar}{m}}d{\bf W}_{t}
\end{equation}

\medskip{}
\noindent and Fokker-Planck equation:
\begin{equation}
\frac{\partial\rho({\bf x},t)}{\partial t}+{\bf \nabla}\cdot\left(\rho({\bf x},t)\frac{{\bf \nabla}S'({\bf x},t)}{m}\right)=\frac{\hbar}{2m}{\bf \nabla}^{2}\rho({\bf x},t),
\end{equation}

\medskip{}
\noindent where the transformed phase is:
\begin{equation}
S'({\bf x},t)=S({\bf x},t)+\frac{\hbar}{2}\log\rho({\bf x},t).\label{eq:30}
\end{equation}

\medskip{}

\medskip{}
\noindent Equation {[}Eq. \ref{eq:31}{]} can now be simplified with
the transformed phase. Using the continuity equation, the first collection
of terms are simplified to:
\begin{align}
\frac{\partial S}{\partial t}-\frac{\hbar}{2m}\nabla^{2}S-\frac{\hbar}{2m}\nabla S\cdot\nabla\log\rho= & \frac{\partial S}{\partial t}+\frac{\hbar}{2}\frac{\partial\log\rho}{\partial t}\label{eq:1-2-1}\\
= & \frac{\partial S'}{\partial t}.
\end{align}
The second collection of terms are simplified to:
\begin{align}
\frac{1}{2m}\nabla S\cdot\nabla S+\frac{\hbar}{2m}\nabla\log\rho\cdot\nabla S+\frac{\hbar^{2}}{8m}\nabla\log\rho\cdot\nabla\log\rho= & \frac{1}{2m}\nabla\left(S+\frac{\hbar}{2}\log\rho\right)\cdot\nabla\left(S+\frac{\hbar}{2}\log\rho\right)\\
= & \frac{1}{2m}\nabla S'\cdot\nabla S'.
\end{align}
The third collection of terms are simplified to:
\begin{align}
\frac{\hbar}{2m}\nabla^{2}S+\frac{\hbar}{4m}\nabla^{2}\log\rho= & \frac{\hbar}{2m}\nabla^{2}\left(S+\frac{\hbar}{2}\log\rho\right)\\
= & \frac{\hbar}{2m}\nabla^{2}S'.
\end{align}
Therefore the quantum Hamilton-Jacobi equation becomes:
\begin{equation}
\frac{\partial S'}{\partial t}+\frac{1}{2m}\nabla S'\cdot\nabla S'+\frac{\hbar}{2m}\nabla^{2}S'+V-\frac{\hbar^{2}}{4m}\nabla^{2}\log\rho-\frac{\hbar^{2}}{4m}\frac{\nabla^{2}\rho}{\rho}=0.
\end{equation}
or equivalently:
\begin{equation}
\frac{\partial S'}{\partial t}+\frac{1}{2m}\nabla S'\cdot\nabla S'+\frac{\hbar}{2m}\nabla^{2}S'+V+2Q=0,\label{eq:4.16}
\end{equation}
where we have substituted the third definition of the quantum potential
{[}Eq. \ref{eq:126}{]} (Appendix \ref{sec:Quantum-potential}). Rearranging
this equation gives:
\begin{equation}
\left[\frac{\partial}{\partial t}+\frac{1}{m}\nabla S'\cdot\nabla+\frac{\hbar}{2m}\nabla^{2}\right]S'=\frac{1}{2m}\nabla S'\cdot\nabla S'-V-2Q.\label{eq:52}
\end{equation}

\medskip{}
\noindent Now using the definition of the retrocausal stochastic
Lagrangian derivative $\frac{D_{*}}{Dt}=\frac{\partial}{\partial t}+\frac{1}{m}\nabla S'\cdot\nabla+\frac{\hbar}{2m}\nabla^{2}$
(Appendix \ref{subsec:Retrocausal-Lagrangian-derivativ}), equation
{[}\ref{eq:52}{]} can be written as:
\begin{align}
\frac{D_{*}S'}{Dt}= & \frac{1}{2m}\nabla S'\cdot\nabla S'-V-2Q.\label{eq:53}
\end{align}

\medskip{}

\subsection{\label{subsec:Removing-non-local-terms}Removing the non-local terms}

While the Lagrangian equation for the transformed phase $S'({\bf x},t)$
{[}Eq. \ref{eq:53}{]} contains non-local quantum potential terms,
using the methods of section \ref{sec:A-conceptual-framework} it
can be shown that they disappear when propagating the transformed
phase. Similar to section \ref{subsec:Elimination-of-the}, we rearrange
the Lagrangian equation to solve for the transformed phase, which
gives a stochastic Hamilton-Jacobi-Bellman equation. Then the stochastic
Hamilton-Jacobi-Bellman equation is recursively expanded. For instance,
expanding twice gives:
\begin{align}
S'({\bf x},t)= & E_{\delta{\bf x}}\left[S'({\bf x}+\delta{\bf x},t+\delta t)\right]-\frac{D_{*}S'({\bf x},t)}{Dt}\delta t\\
= & E_{\delta{\bf x}}\left[E_{\delta{\bf y}}\left[S'({\bf x}+\delta{\bf x}+\delta{\bf y},t+2\delta t)\right]\right]-E_{\delta{\bf x}}\left[\frac{D_{*}S'({\bf x}+\delta{\bf x},t+\delta t)}{Dt}\right]\delta t-\frac{D_{*}S'({\bf x},t)}{Dt}\delta t.\label{eq:54}
\end{align}
We have already shown in section \ref{subsec:Vanishing-expectation-of}
that:
\begin{align}
E_{\delta{\bf x}}\left[\nabla^{2}f({\bf x}+\delta{\bf x},t+\delta t)\right]\approx & 0,
\end{align}
provided that $f({\bf x}+\delta{\bf x},t+\delta t)$ is approximately
constant over the region where the transition probability $P({\bf x}+\delta{\bf x},t+\delta t|{\bf x},t)$
is non-zero. This condition is true for $f({\bf x}+\delta{\bf x},t+\delta t)=\nabla^{2}\log\rho({\bf x}+\delta{\bf x},t+\delta t)$
and consequently:
\begin{equation}
E_{\delta{\bf x}}\left[\frac{\hbar^{2}}{4m}\nabla^{2}\log\rho({\bf x}+\delta{\bf x},t+\delta t)\right]\approx0.
\end{equation}
The same logic is applicable to the $\frac{\hbar^{2}}{4m}\frac{\nabla^{2}\rho}{\rho}$
term of quantum potential in {[}Eq. \ref{eq:53}{]}, but with a slight
difference due to the factor of $\frac{1}{\rho}$ which must be removed
in order to apply the divergence theorem. The expectation is converted
to a sum over particle configurations, and then approximated as integral
over particle transition probabilities: 

\begin{align}
E_{\delta{\bf x}}\left[\frac{\nabla^{2}\rho({\bf x}+\delta{\bf x},t+\delta t)}{\rho({\bf x}+\delta{\bf x},t+\delta t)}\right]= & \sum_{a}\frac{\nabla^{2}\rho(q_{a}(t+\delta t))}{\rho(q_{a}(t+\delta t)))}\\
= & \sum_{{\bf \delta{\bf x}}}P({\bf x}+\delta{\bf x},t+\delta t|{\bf x},t)P({\bf x},t)\frac{\nabla_{{\bf x}+\delta{\bf x}}^{2}\rho({\bf x}+\delta{\bf x},t+\delta t)}{\rho({\bf x}+\delta{\bf x},t+\delta t)}\\
\approx & \int P({\bf y},t+\delta t|{\bf x},t)P({\bf x},t)\frac{\nabla_{{\bf y}}^{2}\rho({\bf y},t+\delta t)}{\rho({\bf y},t+\delta t)}d{\bf y},
\end{align}
where we have replaced the variable ${\bf x}+\delta{\bf x}$ with
${\bf y}$ to simplify the notation. The fraction $\frac{1}{\rho({\bf y},t+\delta t)}$
in the equation can then be removed using Bayes' theorem {[}Eq. \ref{eq:65}{]}
giving:
\begin{align}
E_{\delta{\bf x}}\left[\frac{\nabla^{2}\rho({\bf x}+\delta{\bf x},t+\delta t)}{\rho({\bf x}+\delta{\bf x},t+\delta t)}\right]\approx & \int P({\bf x},t|{\bf y},t+\delta t)\frac{P({\bf y},t+\delta t)}{\rho({\bf y},t+\delta t)}\nabla_{{\bf y}}^{2}\rho({\bf y},t+\delta t)d{\bf y}\\
\approx & \int P({\bf x},t|{\bf y},t+\delta t)\nabla_{{\bf y}}^{2}\rho({\bf y},t+\delta t)d{\bf y},
\end{align}
where Bayes' theorem implies:
\begin{align}
P({\bf y},t+\delta t|{\bf x},t)P({\bf x},t)= & P({\bf x},t|{\bf y},t+\delta t)P({\bf y},t+\delta t),\label{eq:65}
\end{align}
and we have identified that $\frac{P({\bf y},t+\delta t)}{\rho({\bf y},t+\delta t)}=1$.
The inverse product rule and then divergence theorem can now be used
twice to shift the $\nabla_{{\bf y}}^{2}$ operator onto the $P({\bf x},t|{\bf y},t+\delta t)$
term: 
\begin{align}
E_{\delta{\bf x}}\left[\frac{\nabla^{2}\rho({\bf x}+\delta{\bf x},t+\delta t)}{\rho({\bf x}+\delta{\bf x},t+\delta t)}\right]= & \int P({\bf x},t|{\bf y},t+\delta t)\nabla_{{\bf y}}^{2}\rho({\bf y},t+\delta t)d{\bf y}\\
= & \int\rho({\bf y},t+\delta t)\nabla_{{\bf y}}^{2}P({\bf x},t|{\bf y},t+\delta t)d{\bf y}.
\end{align}

\medskip{}
\noindent The probability of a particle originating from position
$({\bf x},t)$ given that it is observed at position (${\bf y},t+\delta t$)
has a finite spatial support around the location ${\bf y=x}$. Therefore
$\rho({\bf y},t+\delta t)$ is approximately constant over this small
region, and it can be taken out of the integral as a constant. 

\begin{align}
E_{\delta{\bf x}}\left[\frac{\nabla^{2}\rho({\bf x}+\delta{\bf x},t+\delta t)}{\rho({\bf x}+\delta{\bf x},t+\delta t)}\right]\approx & \rho({\bf y},t+\delta t)\int\nabla_{{\bf y}}^{2}P({\bf x},t|{\bf y},t+\delta t)d{\bf y}\\
\approx & 0,
\end{align}
where the last line is set to zero using the divergence theorem. Therefore:
\begin{align}
E_{\delta{\bf x}}\left[Q({\bf x}+\delta{\bf x},t+\delta t)\right]= & E_{\delta{\bf x}}\left[-\frac{\hbar^{2}}{8m}\nabla^{2}\log\rho({\bf x}+\delta{\bf x},t+\delta t)\right]+E_{\delta{\bf x}}\left[-\frac{\hbar^{2}}{8m}\frac{\nabla^{2}\rho({\bf x}+\delta{\bf x},t+\delta t)}{\rho({\bf x}+\delta{\bf x},t+\delta t)}\right]\nonumber \\
\approx & 0,\label{eq:69}
\end{align}
and consequently:
\begin{align}
E_{\delta{\bf x}}\left[\frac{DS({\bf x}+\delta{\bf x},t+\delta t)}{Dt}\right]= & E_{\delta{\bf x}}\left[\frac{1}{2m}\nabla S'({\bf x}+\delta{\bf x},t+\delta t)\cdot\nabla S'({\bf x}+\delta{\bf x},t+\delta t)-V({\bf x}+\delta{\bf x},t+\delta t)\right.\nonumber \\
 & \left.-2Q({\bf x}+\delta{\bf x},t+\delta t)\right]\\
\approx & E_{\delta{\bf x}}\left[\frac{1}{2m}\nabla S'({\bf x}+\delta{\bf x},t+\delta t)\cdot\nabla S'({\bf x}+\delta{\bf x},t+\delta t)-V({\bf x}+\delta{\bf x},t+\delta t)\right]\\
\approx & E_{\delta{\bf x}}\left[\mathcal{L}_{c}({\bf x}+\delta{\bf x},t+\delta t)\right],
\end{align}
where $\mathcal{L}_{c}$ is a classical Lagrangian. Therefore, the
equation for $S'({\bf x},t)$ {[}Eq. \ref{eq:54}{]} equals:
\begin{align}
S'({\bf x},t)\approx & E_{\delta{\bf x}}\left[E_{\delta{\bf y}}\left[S({\bf x}+\delta{\bf x}+\delta{\bf y},t+2\delta t)\right]\right]-E_{\delta{\bf x}}\left[\mathcal{L}_{c}({\bf x}+\delta{\bf x},t+\delta t)\right]\delta t-\frac{D_{*}S'({\bf x},t)}{Dt}\delta t,\label{eq:72}
\end{align}
which now contains only classical Lagrangians. The non-local quantum
potential terms also vanish for further expansions of $S'({\bf x},t)$
by the same procedure, and consequently the phase can be expressed
in terms of expectations over classical Lagrangians. The initial non-local
term $\frac{D_{*}S'({\bf x},t)}{Dt}\delta t$ is vanishingly small
and can be ignored.

\section{\label{sec:Local-interpretation-with}Local interpretation with interactions}

\noindent In section \ref{sec:Stochastic-transform-of}, we showed
that the equations for phase propagation in the Lagrangian picture
of stochastic particles do not contain the quantum potential terms.
This section focuses on embedding the resulting equations in three-dimensional
space. There are two main approaches we will discuss. There is a first-order
interpretation, where the particle motion is governed by the first-order
guidance equations and the phase is propagated according to the stochastic
Hamilton-Jacobi equations. Then there is a second-order interpretation
which is analogous to Bohmian mechanics, and describes particle accelerations
instead of phase propagation. 

The first-order interpretation is somewhat problematic to describe
in three-dimensional space, as particle interactions cause the phase
to become a non-separable function in configuration space if understood
in their direct form. Nevertheless, because the phase propagation
depends only on the classical Lagrangian, the non-separability only
arises from interactions via the classical potential, which can be
described by a local flow of information. This is different to the
situation in standard quantum mechanics, where non-separability arises
from both the quantum and classical potentials. Nevertheless, the
second-order Bohmian interpretation may be more natural for describing
the dynamical system in three-dimensional space, as it avoids the
requirement to describe phase propagation directly. 

\subsection{First order phase propagation}

\noindent The equations of phase propagation in the stochastic Hamilton-Jacobi
approach follow from the results of section \ref{sec:Stochastic-transform-of}.
The equation for the recursive expansion of the phase {[}Eq. \ref{eq:72}{]}
was shown to be:
\begin{align}
S'({\bf x},t)\approx & E_{\delta{\bf x}}\left[E_{\delta{\bf y}}\left[S'({\bf x}+\delta{\bf x}+\delta{\bf y},t+2\delta t)\right]\right]-E_{\delta{\bf x}}\left[\mathcal{L}_{c}({\bf x}+\delta{\bf x},t+\delta t)\right],\label{eq:87}
\end{align}
which depends only on classical Lagrangians for the propagation. We
have ignored the $-\frac{DS'({\bf x},t)}{Dt}\delta t$ term in {[}Eq.
\ref{eq:72}{]} as it is the initial piece of the recursive expansion
and is vanishingly small. Although {[}Eq. \ref{eq:87}{]} contains
only the classical Lagrangian, an issue occurs when embedding this
equation in three-dimensional space. If the classical Lagrangian contains
particle interactions, the phase is a non-separable function in configuration
space. Note that the issue arises due to the classical potential only,
not the quantum potential which has been removed. 

Now you may wonder how classical mechanics is able to deal with this
problem. The solution is simple, classical mechanics typically uses
the second-order Newtonian dynamical picture instead of first-order
phase propagation. However, first-order phase propagation is clearly
possible in principle since the information flow is local. Developing
an ontological interpretation does require careful consideration however,
so as to not introduce a large number of physical states to replace
the configuration space description.

To simplify matters, we will make some basic assumptions. Firstly,
we assume the phase propagation is described on a spatial lattice
in three-dimensional space. It is not tenable to assume the phase
is propagated, for example, by a physical ensemble of particle configurations,
which would require an exponentially large set of physical objects
for the description. Secondly, since the phase can be decomposed into
the Lagrangian for free-particle propagation and interaction Lagrangian,
we focus on describing the interaction Lagrangian in three-dimensional
space, as the Lagrangian for free-particle propagation is simple to
understand in three-dimensional space directly. To proceed, the individual
particle only needs to know the gradient of the phase to calculate
its velocity using the first-order de Broglie-Bohm stochastic guidance
equation. Therefore, examine the equation for the gradient of the
phase calculated using {[}Eq. \ref{eq:87}{]}: 
\begin{align}
\nabla_{1}S'({\bf x},t)\approx & \frac{S'({\bf x}+\delta{\bf j}_{1},t)-S'({\bf x},t)}{|\delta{\bf j}_{1}|},\\
\approx & \frac{1}{|\delta{\bf j}_{1}|}E_{\delta{\bf x}}\left[E_{\delta{\bf y}}\left[S'({\bf x}+\delta{\bf x}+\delta{\bf y}+\delta{\bf j},t+2\delta t)-S'({\bf x}+\delta{\bf x}+\delta{\bf y},t+2\delta t)\right]\right]\ldots\nonumber \\
 & \ldots-\frac{1}{|\delta{\bf j}_{1}|}E_{\delta{\bf x}}\left[\mathcal{L}_{c}({\bf x}+\delta{\bf x}+\delta{\bf j}_{1},t+\delta t)-\mathcal{L}_{c}({\bf x}+\delta{\bf x},t+\delta t)\right],\label{eq:5.3}
\end{align}
where $\delta{\bf j}_{1}$ represents a small shift in the position
coordinates of particle 1. Examining the second term of {[}Eq. \ref{eq:5.3}{]}:
\begin{equation}
E_{\delta{\bf x}}\left[\mathcal{L}_{c}({\bf x}+\delta{\bf x}+\delta{\bf j}_{1},t+\delta t)-\mathcal{L}_{c}({\bf x}+\delta{\bf x},t+\delta t)\right]
\end{equation}
 shows how to construct an ontology for the phase propagation in three-dimensional
space. Note that $\mathcal{L}_{c}({\bf y},t)=\mathcal{L}_{c}^{F}({\bf y}_{1},t)+\mathcal{L}_{c}^{F}({\bf y}_{2},t)+\mathcal{L}_{c}^{I}({\bf y}_{1},{\bf y}_{2},t)$,
where $\mathcal{L}_{c}^{F}({\bf y}_{1},t)$ and $\mathcal{L}_{c}^{F}({\bf y}_{2},t)$
are the Lagrangians for free-particle propagation for particle 1 and
2 respectively, and $\mathcal{L}_{c}^{I}({\bf y}_{1},{\bf y}_{2},t)$
is the interaction Lagrangian. Clearly in the differenced quantity
$\mathcal{L}_{c}^{F}({\bf y}_{2},t)$ does not contribute:
\begin{align}
 & E_{\delta{\bf x}}\left[\mathcal{L}_{c}({\bf x}+\delta{\bf x}+\delta{\bf j}_{1},t+\delta t)-\mathcal{L}_{c}({\bf x}+\delta{\bf x},t+\delta t)\right]\\
 & =E_{\delta{\bf x}}\left[\mathcal{L}_{c}^{F}({\bf x}_{1}+\delta{\bf x}_{1}+\delta{\bf j}_{1},t+\delta t)-\mathcal{L}_{c}^{F}({\bf x}_{1}+\delta{\bf x}_{1},t+\delta t)\right]\ldots\nonumber \\
 & \ldots+E_{\delta{\bf x}}\left[\mathcal{L}_{c}^{I}({\bf x}_{1}+\delta{\bf j}_{1},{\bf x}_{2},t+\delta t)-\mathcal{L}_{c}^{I}({\bf x}_{1},{\bf x}_{2},t+\delta t)\right],
\end{align}
The first term depends only upon the coordinates of the first particle.
It can be described, for instance, by propagating the classical free-particle
Lagrangian information on the spatial lattice for particle one, along
stochastic paths. The second part is comprised of interactions via
the classical potential $V$ which make up the interaction Lagrangian.
Imposing the no-local signaling condition implies that only the components
of the interaction Lagrangian which represent the local flow of information
into the single-particle subsystem are available for calculating the
motion of the individual particle. The components of the interaction
Lagrangian which are non-local cancel in the differenced term $\mathcal{L}_{c}^{I}({\bf x}_{1}+\delta{\bf j}_{1},{\bf x}_{2},t+\delta t)-\mathcal{L}_{c}^{I}({\bf x}_{1},{\bf x}_{2},t+\delta t)$.
Locality is ensured in this quantity, because the only way for it
to be non-zero is if the difference in the position of particle one,
i.e. ${\bf x}_{1}+\delta{\bf j}_{1}$ vs. ${\bf x}_{1}$, has caused
a difference in the interaction between particles via the classical
potential along their trajectories. For this to be the case there
has to be a physical connection in the particle's history to the interaction
location, either directly or indirectly by first interacting with
intermediary particles. A local flow of information can propagate
along this path of connection from the interaction location to the
particle's current position. 

\subsection{\label{subsec:Second-order-Newtonian-description}Deterministic Bohmian
mechanics}

Bohmian mechanics describes the motion of particles in terms of their
accelerations rather than via the first-order guidance equation. The
equation of particle acceleration can be derived by taking the deterministic
Lagrangian derivative of the particle velocity:
\begin{align}
\frac{d^{2}{\bf X}}{d^{2}t}=\frac{d}{dt}\left(\frac{d{\bf X}}{dt}\right)= & \frac{1}{m}\frac{d\nabla S({\bf x},t)}{dt}.\label{eq:88}
\end{align}
The Lagrangian derivative $\frac{d\nabla S({\bf x},t)}{dt}$ can now
be found by differentiating the quantum Hamilton-Jacobi equation {[}Eq.
\ref{eq:6-1}{]} with respect to $\nabla$:
\begin{align}
\frac{\partial\nabla S({\bf x},t)}{\partial t}+\frac{1}{m}\nabla S({\bf x},t)\cdot\nabla\nabla S({\bf x},t)+\nabla V({\bf x},t)+\nabla Q({\bf x},t)= & 0.
\end{align}
Therefore:
\begin{align}
\frac{d\nabla S({\bf x},t)}{dt}\equiv & \left[\frac{\partial}{\partial t}+\frac{1}{m}\nabla S({\bf x},t)\cdot\nabla\right]\nabla S({\bf x},t)\\
= & -\nabla\left[V({\bf x},t)+Q({\bf x},t)\right].
\end{align}
Consequently, the Bohmian equation of motion {[}Eq. \ref{eq:88}{]}
is equal to:
\begin{align}
\frac{d^{2}{\bf X}}{d^{2}t}= & -\frac{1}{m}\nabla\left[V({\bf x},t)+Q({\bf x},t)\right].
\end{align}
This is a second-order Newtonian-like equation which describes particle
acceleration given a quantum force $-\nabla Q({\bf x},t)$ and classical
force $-\nabla V({\bf x},t)$. In Bohmian mechanics, it is regarded
as the fundamental equation of motion, while the first-order guidance
equation is interpreted as a dynamical constraint which is satisfied
due to the initial boundary conditions \cite{key-3}. 

\subsubsection{Stochastic case}

In the stochastic case, the second-order Lagrangian derivative can
be defined as an expectation over finite differences: 

\begin{align}
\frac{D_{*}^{2}{\bf X}}{D^{2}t}\equiv & E_{\delta{\bf x}}\left[\frac{1}{\delta t}\left(\frac{d{\bf X}(t+\delta t)}{\delta t}-\frac{d{\bf X}(t)}{\delta t}\right)\right]\\
= & E_{\delta{\bf x}}\left[\frac{\nabla S'({\bf x}+\delta{\bf x},t+\delta t)\delta t/m+\sqrt{\hbar/m}d{\bf W}_{t+\delta t}-\nabla S'({\bf x},t)\delta t/m-\sqrt{\hbar/m}d{\bf W}_{t}}{\delta t^{2}}\right],
\end{align}
where we have substituted $d{\bf X}=\frac{1}{m}\nabla S'({\bf x},t)\delta t+\sqrt{\frac{\hbar}{m}}d{\bf W}_{t}$
which is the guidance equation for stochastic particles. Since the
expectation of a Wiener processes is zero, the $d{\bf W}_{t}$ terms
are removed, leaving:
\begin{align}
\frac{D_{*}^{2}{\bf X}}{D^{2}t}= & \frac{1}{m}E_{\delta{\bf x}}\left[\frac{\nabla S'({\bf x}+\delta{\bf x},t+\delta t)-\nabla S'({\bf x},t)}{\delta t}\right]\\
= & \frac{1}{m}E_{\delta{\bf x}}\left[\frac{D_{*}\nabla S'({\bf x},t)}{Dt}\right],
\end{align}
which is an equation analogous to {[}Eq. \ref{eq:88}{]}. The result
for $\frac{D_{*}\nabla S({\bf x},t)}{Dt}$ can now be found by differentiating
the transformed Hamilton-Jacobi equation {[}Eq. \ref{eq:4.16}{]}
with respect to $\nabla$: 
\begin{equation}
\frac{\partial\nabla S'({\bf x},t)}{\partial t}+\frac{1}{m}{\bf \nabla}S'({\bf x},t)\cdot{\bf \nabla}\left(\nabla S'({\bf x},t)\right)+\frac{\hbar}{2m}\nabla^{2}\nabla S'({\bf x},t)+\nabla V({\bf x},t)+2\nabla Q=0.\label{eq:26-1}
\end{equation}
Collect terms on left-hand side to form the stochastic Lagrangian
derivative:
\begin{equation}
\left[\frac{\partial}{\partial t}+\frac{1}{m}{\bf \nabla}S'({\bf x},t)\cdot{\bf \nabla}+\frac{\hbar}{2m}\nabla^{2}\right]\nabla S'({\bf x},t)=-\nabla V({\bf x},t)-2\nabla Q.\label{eq:26-1-1}
\end{equation}
Consequently the stochastic Lagrangian derivative of $\nabla S'({\bf x},t)$
is:
\begin{align}
\frac{D_{*}\nabla S'({\bf x},t)}{Dt}= & -\nabla V({\bf x},t)-2\nabla Q({\bf x},t).\label{eq:5.19}
\end{align}
Recognising as was shown in section \ref{subsec:Removing-non-local-terms}
that $E_{\delta{\bf x}}\left[Q({\bf x}+\delta{\bf x},t+\delta t)\right]=0$,
we can also prove: 
\begin{align}
E_{\delta{\bf x}}\left[\nabla Q({\bf x}+\delta{\bf x},t+\delta t)\right]\propto & \sum_{\delta{\bf x}}\nabla Q({\bf x}+\delta{\bf x},t+\delta t)\\
\propto & \nabla\sum_{\delta{\bf x}}Q({\bf x}+\delta{\bf x},t+\delta t)\\
\propto & \nabla E_{\delta{\bf x}}\left[Q({\bf x}+\delta{\bf x},t+\delta t)\right]\approx0.
\end{align}
Therefore the equation for $\frac{D_{*}^{2}{\bf X}}{D^{2}t}$ is:
\begin{align}
\frac{D_{*}^{2}{\bf X}}{D^{2}t}\equiv & \frac{1}{m}E_{\delta{\bf x}}\left[\left(\frac{D_{*}\nabla S'({\bf x}+\delta{\bf x},t+\delta t)}{Dt}\right)\right]\\
= & \frac{1}{m}E_{\delta{\bf x}}\left[-\nabla V({\bf x}+\delta{\bf x},t+\delta t)\right]-\frac{2}{m}E_{\delta{\bf x}}\left[\nabla Q({\bf x}+\delta{\bf x},t+\delta t)\right]\\
\approx & \frac{1}{m}E_{\delta{\bf x}}\left[-\nabla V({\bf x}+\delta{\bf x},t+\delta t)\right].\label{eq:91-1}
\end{align}
Although this equation appears entirely classical and analogous to
Newton's second-law, it may need to be solved as a stochastic Hamilton-Jacobi-Bellman
equation due to the expectation over particle trajectories.

\subsubsection{\label{subsec:Interpretation-in-three-dimensio}Interpretation in
three-dimensional space}

To relate the stochastic equations of motion to particle propagation,
equation {[}Eq. \ref{eq:91-1}{]} can be rearranged in terms of the
particle velocity $\frac{D{\bf X}(t)}{Dt}$: 

\begin{equation}
\frac{D_{*}{\bf X}(t)}{Dt}=E_{\delta{\bf x}}\left[\frac{D_{*}{\bf X}(t+\delta t)}{Dt}+\frac{1}{m}\nabla V({\bf x}+\delta{\bf x},t+\delta t)\delta t\right].\label{eq:89}
\end{equation}
This new equation can be solved for the particle position by substituting
the definition of particle velocity $\frac{D{\bf X}({\bf x},t)}{Dt}\equiv E_{\delta{\bf x}}\left[\delta t^{-1}\left({\bf X}(t+\delta t)-{\bf X}(t)\right)\right]$:

\begin{equation}
E_{\delta{\bf x}}\left[\delta t^{-1}\left({\bf X}(t+\delta t)-{\bf X}(t)\right)\right]=E_{\delta{\bf x}}\left[\frac{D_{*}{\bf X}(t+\delta t)}{Dt}+\frac{1}{m}\nabla V({\bf x}+\delta{\bf x},t+\delta t)\delta t\right]
\end{equation}

\begin{align}
\Rightarrow{\bf X}(t)= & E_{\delta{\bf x}}\left[{\bf X}(t+\delta t)-\frac{D_{*}{\bf X}(t+\delta t)}{Dt}\delta t-\frac{1}{m}\nabla V({\bf x}+\delta{\bf x},t+\delta t)\delta t^{2}\right].\label{eq:91}
\end{align}
To obtain a three-dimensional space description of {[}Eq. \ref{eq:91}{]},
we firstly write this equation for the particle configuration ${\bf X}=(X_{1},...,X_{N})$
in terms of equations for the individual particles: 
\begin{align}
X_{i}(t)= & E_{\delta{\bf x}}\left[X_{i}(t+\delta t)-\frac{D_{*}X_{i}(t+\delta t)}{Dt}\delta t-\frac{1}{m}\nabla_{i}V({\bf x}+\delta{\bf x},t+\delta t)\delta t^{2}\right].\label{eq:92}
\end{align}
An interpretation of this equation {[}Eq. \ref{eq:92}{]} is that
it is analogous to a stochastic Hamilton-Jacobi-Bellman equation.
In this stochastic Hamilton-Jacobi-Bellman interpretation of the dynamics,
$X_{i}(t)$ is now regarded not a single particle, but can be thought
of as being comprised of a sub-ensemble of particles, or alternatively
as information propagating on a three-dimensional space lattice in
the stochastic dynamic programming picture. The time-evolution due
to {[}Eq. \ref{eq:92}{]} can be understood in matrix form. Let:
\begin{equation}
\vec{X}_{i}(t)=\left(\begin{array}{ccc}
X_{i1}(t), & \ldots, & X_{iN}(t)\end{array}\right)^{T},
\end{equation}
\begin{equation}
\vec{v}_{i}(t)=\left(\begin{array}{c}
\frac{DX_{i1}(t)}{Dt}\\
\vdots\\
\frac{DX_{iN}(t)}{Dt}
\end{array}\right),
\end{equation}

\begin{equation}
\vec{a}_{i}(t)=\left(\begin{array}{c}
-\frac{1}{m}\nabla_{i1}V({\bf x},t)\delta(x_{1}=X_{1}(t))\\
\vdots\\
-\frac{1}{m}\nabla_{iN}V({\bf x},t)\delta(x_{N}=X_{N}(t))
\end{array}\right),
\end{equation}
be the set of positions, velocities and accelerations for particles
of the sub-ensemble for particle $i$ respectively. The subscripts
$i$ denote the individual particles, while the subscripts $j\in\mathbb{N}\setminus\{0\}$
denote the elements of the sub-ensemble for that particle. Using this
vector notation, {[}Eq. \ref{eq:92}{]} is equivalent to the matrix
equation:
\begin{equation}
\vec{X}_{i}(t)=T^{-1}\left(\vec{X}_{i}(t+\delta t)-\vec{v}_{i}(t+\delta t)\delta t+\vec{a}_{i}(t+\delta t)\delta t^{2}\right),\label{eq:96-1}
\end{equation}
where $T$ is the stochastic matrix of transition probabilities for
elements of the sub-ensemble to move from $X_{ij}(t)$ to $X_{ij}(t+\delta t)+\delta x_{ij}$.
Updating the particle positions involves applying the transition probability
matrix $T^{-1}$ to the vector $\vec{X}_{i}(t+\delta t)-\vec{v}_{i}(t+\delta t)\delta t+\vec{a}_{i}(t+\delta t)\delta t^{2}$.
Note that the particle velocities are also updated according to the
same transition probability matrix, and we have due to {[}Eq. \ref{eq:89}{]}:
\begin{equation}
\vec{v}_{i}(t)=T^{-1}\left[\vec{v}_{i}(t+\delta t)-\vec{a}_{i}\left(t+\delta t\right)\delta t\right].\label{eq:89-1}
\end{equation}
Because the particle positions and velocities are updated simultaneously
using the same transition probability matrix $T$, it avoids the problem
of closed causal loops which would arise if performing these updates
sequentially. Closed causal loops potentially could occur because
specifying the particle velocities determines the particle positions,
which gives rise to circular reasoning when attempting to update the
particle position directly. 

Together, equations {[}Eq. \ref{eq:96-1}{]} and {[}Eq. \ref{eq:89-1}{]}
specify the particle dynamics. These equations are both expressed
in terms of the local quantities $\vec{X}_{i}(t+\delta t)$, $\vec{v}_{i}(t+\delta t)$
and $\vec{a}_{i}\left(t+\delta t\right)$, which are accessible to
particle $i$. Therefore the propagation of the particle sub-ensemble
$\vec{X}_{i}(t)$ can be described locally in three-dimensional space.
Since each of the particle sub-ensembles are propagated in three-dimensional
space, the ensemble of particle configurations $\hat{{\bf X}}(t)$
is also propagated in three dimensional space, where $\hat{{\bf X}}(t)$
is defined as:
\begin{equation}
\vec{{\bf X}}(t)\equiv\left(\begin{array}{c}
{\bf X}_{1}(t)\\
\vdots\\
{\bf X}_{N}(t)
\end{array}\right),\label{eq:101}
\end{equation}
with each of the ${\bf X}_{i}(t)$ denoting particle configurations
${\bf X}_{j}=(X_{1j},\cdots,X_{Nj})$. The set of particle configurations
is a rearrangement of the elements of the particle sub-ensembles,
$\vec{X}_{i}(t)=\left(\begin{array}{ccc}
X_{i1}(t), & \ldots, & X_{iN}(t)\end{array}\right)^{T}$. Note the subscript $j$ is in the second position for the particle
configurations compared to the subscript $i$ in the first position
for the particle sub-ensembles. This indicates that the particle configuration
and particle sub-ensembles are different ways to group the same total
set of elements $\{X_{ij}|\ i\in[1,N],j\in[1,N]\}$, i.e. the particle
configuration groups by column, and the sub-ensemble groups by row. 

The presence of two different groupings indicates how entanglement
correlations could occur. Despite propagating locally, individual
elements $X_{ij}$ of the sub-ensembles become correlated due to the
dynamics. The set of elements remains separable by the row groupings.
However, as the elements become correlated, the set of particle configurations
$\{{\bf X}_{j}|\ j\in[1,N]\},$ which are grouped by columns, become
non-separable. Then it becomes impossible to describe $\hat{{\bf X}}(t)$
as tuple of three-dimensional space particle ensembles such as $\hat{{\bf X}}(t)=\left(\vec{X}_{1}(t),\cdots,\vec{X}_{N}(t)\right)$.
Instead, $\hat{{\bf X}}(t)$ must be written in terms of non-separable
configuration space elements $\hat{{\bf X}}(t)=\left({\bf X}_{1}(t),\cdots,{\bf X}_{N}(t)\right)$.
This discussion highlights a novel mechanism by which entanglement
correlations can be explained using retrocausal local hidden variables.
If the fundamental statistical ensemble has two-levels of hierarchy,
for example $X_{ij}$ which has indices $i$ and $j$, then it is
possible for there to be a separable way to group the elements (e.g.
by index $i$) and a non-separable way (e.g. by index $j$). If what
we see in measurements is the non-separable view, then the elements
of the statistical ensemble can be local while simultaneously being
non-separable in our view of the world. 

\section{\label{sec:From particles to policies}Physics in the language of
dynamic programming }

\noindent In this section, we map key concepts in dynamic programming
to the physics of the de Broglie-Bohm interpretation. Doing so provides
new perspectives on quantum mechanics. Of particular interest is the
concept of policies in stochastic dynamic programming, which may either
replace or complement the concept of particle trajectories in the
de Broglie-Bohm interpretation. Another point of interest is the application
of discount factors in Bellman equations, which may cause an intrinsic
form of decoherence through non-unitary time evolution. 

\subsection{Tabulation vs. memoisation }

Dynamic programming has two main approaches, tabulation which is computed
in a bottom-up manner using value functions, and memoisation which
is computed in a top-down manner using recursion. Although memoisation
is interesting to consider, it does not appear to be viable for the
construction of a physical theory. The main problem with the memoisation
approach is that it generates a large recursive call stack which must
be stored in memory. This memory overhead is not feasible in the context
of a physical dynamical theory, which is more simply described by
the tabulation (i.e. value function) approach. Dynamic programming
using the tabulation approach directly relates to the concept of physical
fields being updated through equations of motion. In the case of the
de Broglie-Bohm interpretation for example, the value function can
be regarded as the quantum phase. Updating the value function is equivalent
to updating the quantum phase using the quantum Hamilton-Jacobi equations. 

\subsection{Curse of dimensionality }

The problem of the configuration space description of the wavefunction
in quantum mechanics is similar to the curse of dimensionality in
dynamic programming. In both cases, the problem is that state space
of the problem is exponentially large. There may be approximation
techniques that can be borrowed from the discipline of dynamic programming
that reduce the state space, which can be applied to the quantum mechanical
description. 

\subsection{Discount factor }

Many dynamic programming problems include a discount factor in the
Bellman equations. It is possible that quantum theory when expressed
as a dynamic program also has a discount factor, which would prevent
entanglement effects from propagating over large time horizons. The
discount factor would act as a source of an intrinsic decoherence-like
effect, which couldn't be removed even in a perfect experimental setup.
If observed in present-day experiments, the effect of a discount factor
may be misattributed to decoherence. Furthermore, the effect of a
discount factor may be masked by decoherence if the effect is small.
However it may become noticeable as decoherence is suppressed through
better experimental techniques. Observable signatures of a discount
factor may be decoherence-like effects occurring in an orderly manner
e.g. entanglement information decaying at a uniform rate across many
different experimental setups. This may indicate the presence of a
more fundamental non-unitary behavior than from interaction driven
decoherence. 

This is important because quantum computers are providing increasingly
advanced experimental tests of the high-dimensional mathematical description
of the Copenhagen interpretation. We are exploring unknown territory
as the number of qubits and decoherence timescales increase. As evidence
is gathered through further experimental realizations of quantum computing,
it is highly possible that limitations to the standard quantum description
of unitary time evolution become uncovered. For instance, nature may
provide an information firewall; for instance a fundamental source
of decoherence, as in the hypothetical discount factor; which prevents
certain computational algorithms from being completed. 

Indeed, it is difficult to imagine that nature can store all possible
worlds described by the unitary time-evolution of the wavefunction.
A discount factor would prevent the state space of the universe from
becoming exponentially large. It would act as intrinsic form of wavefunction
collapse, independent from the notions of an observer or measurement
process. 

\subsection{From particles to policies}

The deterministic de Broglie-Bohm interpretation is analogous to deterministic
dynamic programming problems. Here, the particle trajectory is equivalent
to the computational solution of the dynamic programming problem,
while the quantum phase is equivalent to the value function. Note
that the gradient of a function is a vector in the direction of the
local maximum. This produces a correspondence to dynamic programming
through the Bellman equations which describe maximisation or minimisation
problems. In stochastic dynamic programming however, there is not
a single unique computational path chosen. Instead, what is optimized
for is the policy function, which records the optimal action to take
in any given state. Therefore, we may need to consider the de Broglie-Bohm
interpretation not in terms of particles, but in terms of policy functions
if considered as a stochastic dynamic program. One interpretation
is that the policy function is analogous to the gradient of the phase
$\nabla S$ in the de Broglie-Bohm interpretation, while the value
function remains the phase $S$. Alternatively, in the second-order
Bohmian interpretation, the policy function may be regarded as the
gradient of the quantum potential $-\nabla Q$, while the value function
is the quantum potential. 

\subsection{Value iteration vs. policy iteration }

Having introduced the concept of dynamic programming policy functions
in the de Broglie-Bohm interpretation, an opportunity to describe
theory in terms of policy iteration instead of value iteration arises.
Value iteration and policy iteration are both valid computational
methods for dynamic programming problems in the tabulation approach,
in both the deterministic and stochastic cases. The stochastic de
Broglie-Bohm interpretation presented in previous sections has been
described using a value iteration ontology. The general approach of
value iteration is to simultaneously propagate the value function
(i.e. the quantum phase) and the policy function (i.e. the de Broglie-Bohm
particle trajectories or $\nabla S$) in the background of the time
coordinate $t$. 

The approach of policy iteration is quite different from this picture.
In policy iteration, an arbitrary value function and policy are chosen.
Multiple rounds of policy improvement and policy evaluation are conducted
until the value function converges. This algorithm is guaranteed to
converge to the optimal value function and optimal policy after a
sufficient number of iterations. The benefit of policy iteration is
that while performing policy evaluation is computationally cheap,
extracting the optimal policy for each state given a value function
is computationally expensive, especially if the state space is large.
Policy iteration can bypass many calculations of this computationally
expensive step by applying approximations, for instance making stochastic
improvements to the policy. A randomized accept-reject algorithm can
be proposed to update the policy function for example. How this may
work is that random changes to the policy for each state can be proposed,
and if the proposed change improves the value function locally, then
update the policy, otherwise discard the change. Why this may be beneficial
is that the algorithm can operate on the local states of the policy
function and value function simultaneously. The picture is not a dynamical
time-evolution necessarily, but a local process occurring in a Block
universe. Over many iterations the policy and value function will
evolve to contain entanglement-like correlations. 

Policy iteration conducted in this manner highlights a potential mechanism
by which the configuration space nature of the quantum phase can be
explained. Firstly, we assume the value function and policy are updated
stochastically using policy iteration. Therefore, we hypothesize that
the ordinary quantum phase in configuration space $S({\bf x}_{1},{\bf x}_{2},t)$
is actually the statistical average over the value functions calculated
through the stochastic policy iteration procedure, where hidden variables
$\lambda$ identify each individual world of the statistical ensemble.
In other words, we suppose that: 
\begin{equation}
S({\bf x}_{1},{\bf x}_{2},t)=\frac{1}{N}\sum_{\lambda=1}^{N}S({\bf x}_{1},{\bf x}_{2},t|\lambda)P(\lambda),
\end{equation}
where the value function $S({\bf x}_{1},{\bf x}_{2},t|\lambda)$ is
that computed for an individual world. Now we suppose that {\small{}$S({\bf x}_{1},{\bf x}_{2},t|\lambda)=S({\bf x}_{1},t|\lambda)+S({\bf x}_{2},t|\lambda)$}
is separable. Therefore, the information can be written alternatively
a tuple {\small{}$\left(S({\bf x}_{1},t|\lambda),S({\bf x}_{2},t|\lambda)\right)$}.
Similarly, the policy for a given world can be written as a tuple{\small{}
$(\nabla_{1}S({\bf x},t|\lambda),\nabla_{2}S({\bf x},t|\lambda))$}.
Note that this is a local hidden variable theory for quantum mechanics.
It is enabled by the retrocausal nature of the policy iteration algorithm. 

When the particle is freely propagating in the absence of interactions,
the policy improvement step is based upon the Lagrangian for the free-particle,
which contains no interaction term, and no quantum potential term
in the stochastic dynamic programming picture we have presented in
earlier sections of this paper. When there is an interaction between
particles at a particular location $({\bf x}_{1},{\bf x}_{2},t)$,
the two value functions $\left(S({\bf x}_{1},t|\lambda),S({\bf x}_{2},t|\lambda)\right)$,
and the two policies $(\nabla_{1}S({\bf x},t|\lambda),\nabla_{2}S({\bf x},t|\lambda))$
become correlated. This correlation eventually propagates from the
location $({\bf x}_{1},{\bf x}_{2},t)$ throughout the entire policy
and value function through multiple rounds of the policy iteration
procedure. 

This description is very similar to the general approach of Bohmian
mechanics toward the configuration space probability density. In Bohmian
mechanics, it is assumed there is a single particle configuration
chosen from the statistical ensemble of particle configurations. The
probability density of quantum mechanics is regarded as the ensemble
average over correlated particle configurations. Now we are suggesting
that it is not just particles configurations which are elements of
a statistical ensemble, the value function $\left(S({\bf x}_{1},t|\lambda),S({\bf x}_{2},t|\lambda)\right)$
and policy function $(\nabla_{1}S({\bf x},t|\lambda),\nabla_{2}S({\bf x},t|\lambda))$
are also elements. Not only does the statistical average of particle
configurations become non-separable, the statistical average of the
policy and value function also become non-separable in this description.
Therefore, it represents a more general version of the de Broglie-Bohm
scheme. The meaning of this new form of correlation requires further
elucidation. 

\section{\label{sec:Alternative-transformation-of}Retrocausal transformation
of the quantum Hamilton-Jacobi equation}

\noindent In this section, we present an alternative transformation
of the quantum Hamilton-Jacobi equation based on the retrocausal transformation
of the phase $S_{*}'=S-\frac{\hbar}{2}\log\rho$. This is in contrast
to the previous forward-causal transformation of the phase $S'=S+\frac{\hbar}{2}\log\rho$. 

\subsection{Retrocausal stochastic transformation }

We assume the de Broglie-Bohm particle configuration undergoes a stochastic
motion with guidance equation:
\begin{equation}
d{\bf X}(t)=\frac{{\bf \nabla}S_{*}'({\bf x},t)}{m}dt+\sqrt{\frac{k}{m}}d{\bf W}_{t},\label{eq:6-2}
\end{equation}
where $S_{*}'$ is defined as $S'_{*}=S-\frac{\hbar}{2}\log\rho$.
Particles with this guidance equation satisfy the retrocausal Fokker-Planck
equation:
\begin{equation}
\frac{\partial\rho({\bf x},t)}{\partial t}+{\bf \nabla}\cdot\left(\rho({\bf x},t)\frac{{\bf \nabla}S_{*}'({\bf x},t)}{m}\right)=-\frac{\hbar}{2m}{\bf \nabla}^{2}\rho({\bf x},t),\label{eq:12-1}
\end{equation}
Now we also transform the quantum Hamilton-Jacobi equation into the
picture of the phase $S_{*}'$. Adding and subtracting several terms
to the quantum Hamilton-Jacobi equation gives {[}Eq. \ref{eq:4.4}{]}
reproduced here:
\begin{align}
\frac{\partial S_{*}}{\partial t}+\frac{1}{2m}\nabla S_{*}\cdot\nabla S_{*}+V+\left[\frac{\hbar}{2m}\nabla^{2}S_{*}-\frac{\hbar}{2m}\nabla^{2}S_{*}\right]+\left[\frac{\hbar^{2}}{4m}\nabla^{2}\log\rho-\frac{\hbar^{2}}{4m}\nabla^{2}\log\rho\right]\nonumber \\
+\left[\frac{\hbar}{2m}\nabla\log\rho\cdot\nabla S_{*}-\frac{\hbar}{2m}\nabla\log\rho\cdot\nabla S_{*}\right]+\frac{\hbar^{2}}{8m}\nabla\log\rho\cdot\nabla\log\rho-\frac{\hbar^{2}}{4m}\frac{\nabla^{2}\rho}{\rho} & =0.
\end{align}
Which is rearranged into an alternative form:
\begin{align}
\frac{\partial S_{*}}{\partial t}+\left[\frac{\hbar}{2m}\nabla^{2}S_{*}+\frac{\hbar}{2m}\nabla\log\rho\cdot\nabla S_{*}\right]+\left[\frac{1}{2m}\nabla S_{*}\cdot\nabla S_{*}-\frac{\hbar}{2m}\nabla\log\rho\cdot\nabla S_{*}+\frac{\hbar^{2}}{8m}\nabla\log\rho\cdot\nabla\log\rho\right]+\nonumber \\
\left[-\frac{\hbar}{2m}\nabla^{2}S_{*}+\frac{\hbar^{2}}{4m}\nabla^{2}\log\rho\right]+V-\frac{\hbar^{2}}{4m}\nabla^{2}\log\rho-\frac{\hbar^{2}}{4m}\frac{\nabla^{2}\rho}{\rho} & =0.
\end{align}
The first collection of terms are:
\begin{equation}
\left[\frac{\hbar}{2m}\nabla^{2}S_{*}+\frac{\hbar}{2m}\nabla\log\rho\cdot\nabla S_{*}\right]=-\frac{\partial\frac{\hbar}{2}\log\rho}{\partial t}.
\end{equation}
The second collection of terms are:
\begin{align}
\frac{1}{2m}\nabla S_{*}\cdot\nabla S_{*}-\frac{\hbar}{2m}\nabla\log\rho\cdot\nabla S_{*}+\frac{\hbar^{2}}{8m}\nabla\log\rho\cdot\nabla\log\rho= & \frac{1}{2m}\nabla S_{*}'\cdot\nabla S_{*}'.
\end{align}
The third collection of terms are:
\begin{align*}
\left[-\frac{\hbar}{2m}\nabla^{2}S_{*}+\frac{\hbar^{2}}{4m}\nabla^{2}\log\rho\right]= & -\frac{\hbar}{2m}\nabla^{2}S_{*}'.
\end{align*}
And the quantum potential term is again identified using {[}Eq. \ref{eq:126}{]}
(Appendix \ref{sec:Quantum-potential}): 
\begin{align}
-\frac{\hbar^{2}}{4m}\nabla^{2}\log\rho-\frac{\hbar^{2}}{4m}\frac{\nabla^{2}\rho}{\rho}= & 2Q.
\end{align}
Therefore:
\begin{align}
\frac{\partial S_{*}'}{\partial t}+\frac{1}{2m}\nabla S_{*}'\cdot\nabla S_{*}' & -\frac{\hbar}{2m}\nabla^{2}S_{*}'+V+2Q=0.
\end{align}
This transformed quantum Hamilton-Jacobi equation can be expressed
in terms of the forward-causal stochastic Lagrangian derivative as:
\begin{align}
\frac{DS'}{Dt}= & \frac{1}{2m}\nabla S_{*}'\cdot\nabla S_{*}'-2Q-V,
\end{align}
where the forward-causal stochastic Lagrangian derivative (Appendix
\ref{subsec:Forward-causal-Lagrangian-deriva}) is defined as:
\begin{align}
\frac{D}{Dt}= & \frac{\partial}{\partial t}+\frac{1}{m}\nabla S_{*}'\cdot\nabla-\frac{\hbar}{2m}\nabla^{2}S_{*}'.
\end{align}
So we see that if the de Broglie-Bohm particles are forward-causal,
the transformed quantum Hamilton-Jacobi equations are retrocausal
as shown in section \ref{sec:Stochastic-transform-of}. However, if
particles are retrocausal, the transformed quantum Hamilton-Jacobi
equations are forward-causal as shown in this section. In both cases
the phase appears to propagate in the opposite temporal direction
of the particle. 

\subsection{Boundary conditions and absence of fine-tuning }

In section \ref{sec:Local-interpretation-with}, we have elected to
use the solution where the particle is forward-causal and the phase
propagation is retrocausal. There is a good reason for making this
choice, which can be seen by analogy to dynamic programming. In dynamic
programming, there are two possible ways to compute the value function.
The first way is to calculate it from the initial row to the final
row of the computational grid, taking the initial boundary condition
to be a row of zero values. Once the value function has been computed,
the optimal solution is determined using backward induction from the
final row to the initial row. This method is analogous to the case
of forward-causal phase propagation with a retrocausal particle configuration.
It is problematic however, because the particle configuration is required
to be fine-tuned in the initial conditions and have knowledge of future
values. 

The second approach to dynamic programming is to compute the value
function from the final row to the initial row of the computational
grid, taking the final boundary conditions to be a row of zero values.
In this case, the computational solution is calculated in the forward
direction of the grid. This approach is analogous to having the quantum
phase propagating retrocausally, with the particle propagating in
the forward direction of time. There is no fine-tuning of the particle
distribution in this second-choice of interpretation, as the particle
distribution begins in a generic state with no information about the
future encoded into the positions. The particle distribution can evolve
into a range of compatible final outcomes, and therefore does not
have to be precisely specified. Indeed, the dynamics can be stochastic
or even grossly misspecified, as in the case of dynamical relaxation
to quantum equilibrium \cite{key-7}, and still recover the quantum
predictions.

Curiously, the second choice is very different to how the quantum
phase is usually imagined. It is typically assumed that complexity
in the quantum phase grows with time. However, in making the second
choice of retrocausal phase propagation, it is evident that complexity
in the phase decreases with time, as the phase progresses to a final
boundary condition of zero. Nevertheless, the second choice is entirely
compatible with the quantum Hamilton-Jacobi equations, and the two
situations are symmetrical. The symmetry is akin to the symmetry between
the forward and backward methods of computing the value function in
dynamic programming; both of which are acceptable solutions to the
problem. Perhaps then, the quantum phase propagation has been misunderstood,
and the wrong boundary condition is typically assumed, leading to
the paradox of non-locality. 

In dynamic programming, the symmetry arises because the extremal path
computed in one direction is also the extremal path computed in the
opposite direction. In quantum mechanics, it is known through the
Feynman path integral interpretation that the quantum trajectories
follow the principle of least action. The principle of least action
is similarly symmetric; the total action along the trajectory computed
from final to initial is the same as the total action computed from
initial to final. If the particle takes the path which minimises the
action, this can be understood as minimising from final to initial,
or alternatively minimising from initial to final. Therefore, the
principle of least action imposes a symmetry where the final boundary
condition for the phase can be set to zero instead of the initial
boundary conditions. 

\section{\label{sec:Stochastic-transformations-of}Stochastic transformations
of the classical Hamilton-Jacobi equation}

\noindent In section \ref{sec:Stochastic-transform-of}, we found
that the quantum phase can undergo a forward-causal transformation
$S^{'}=S+\frac{\hbar}{2}\log\rho$, whereby the quantum Hamilton-Jacobi
equation becomes: 
\begin{align}
\frac{D_{*}S'}{Dt}= & \frac{1}{2m}\nabla S'\cdot\nabla S'-2Q-V,
\end{align}
where $\frac{D_{*}}{Dt}=\frac{\partial}{\partial t}+\frac{1}{m}\nabla S'\cdot\nabla+\frac{\hbar}{2m}\nabla^{2}$
is the retrocausal stochastic Lagrangian derivative. In section \ref{sec:Alternative-transformation-of}
we found alternatively that if the phase undergoes a retrocausal transformation
$S_{*}^{'}=S-\frac{\hbar}{2}\log\rho,$ the quantum Hamilton-Jacobi
equation becomes: 
\begin{align}
\frac{DS_{*}'}{Dt}= & \frac{1}{2m}\nabla S_{*}'\cdot\nabla S_{*}'-2Q-V,
\end{align}
where $\frac{D}{Dt}=\frac{\partial}{\partial t}+\frac{1}{m}\nabla S_{*}'\cdot\nabla+\frac{\hbar}{2m}\nabla^{2}$
is the forward-causal stochastic Lagrangian derivative. These transformations
were performed upon the quantum Hamilton-Jacobi equation, taking the
equations of quantum mechanics as the precept. However, it is possible
to apply the same transformations to the classical Hamilton-Jacobi
equation. When applied to the classical system, these transformations
give rise to Nelson's stochastic mechanics. 

\subsection{Classical transformation 1}

Take the classical Hamilton-Jacobi equation: 
\begin{equation}
\frac{\partial S_{*}'}{\partial t}+\frac{1}{2m}\nabla S_{*}'\cdot\nabla S_{*}'+V=0,\label{eq:7.5}
\end{equation}
and retrocausal classical diffusion process, given by the retrocausal
Fokker-Planck equation: 
\begin{equation}
\frac{\partial\rho}{\partial t}+\nabla\cdot\left(\rho\frac{\nabla S_{*}'}{m}\right)=-\frac{1}{2m}\nabla^{2}\rho.\label{eq:7.6}
\end{equation}
Using similar steps as in section \ref{sec:Stochastic-transform-of},
with $S=S_{*}'+\frac{\hbar}{2}\log\rho$ the classical Hamilton-Jacobi
equation transforms to a quantum-like Hamilton-Jacobi equation with
retrocausal stochastic Lagrangian derivative: 
\begin{align}
\frac{D_{*}S}{Dt}= & \frac{1}{2m}\nabla S\cdot\nabla S-Q-V,
\end{align}
and gradient of the Hamilton-Jacobi equation: 
\begin{align}
\frac{D_{*}\frac{1}{m}\nabla S}{Dt}= & -\frac{1}{m}\nabla Q-\frac{1}{m}\nabla V,
\end{align}
which is a Bohmian-like equation of motion. Meanwhile, the Fokker-Planck
equation {[}Eq. \ref{eq:7.6}{]} transforms to a deterministic continuity
equation for the phase $S$: 
\begin{equation}
\frac{\partial\rho}{\partial t}+\nabla\cdot\left(\rho\frac{\nabla S}{m}\right)=0.
\end{equation}

\subsection{Classical transformation 2}

Now take the classical Hamilton-Jacobi equation: 
\begin{equation}
\frac{\partial S'}{\partial t}+\frac{1}{2m}\nabla S'\cdot\nabla S'+V=0,\label{eq:7.1}
\end{equation}
and forward-causal classical diffusion process, given by the forward-causal
Fokker-Planck equation: 
\begin{equation}
\frac{\partial\rho}{\partial t}+\nabla\cdot\left(\rho\frac{\nabla S'}{m}\right)=\frac{1}{2m}\nabla^{2}\rho.\label{eq:7.2}
\end{equation}
Using similar steps as in section \ref{sec:Alternative-transformation-of},
with $S=S'-\frac{\hbar}{2}\log\rho$ the classical Hamilton-Jacobi
equation transforms to a quantum-like Hamilton-Jacobi equation with
forward-causal stochastic Lagrangian derivative: 
\begin{align}
\frac{DS}{Dt}= & \frac{1}{2m}\nabla S\cdot\nabla S-Q-V,
\end{align}
and gradient of the Hamilton-Jacobi equation: 
\begin{align}
\frac{D\frac{1}{m}\nabla S}{Dt}= & -\frac{1}{m}\nabla Q-\frac{1}{m}\nabla V.
\end{align}
The Fokker-Planck equation {[}Eq. \ref{eq:7.2}{]} transforms to a
deterministic continuity equation for the phase $S$: 
\begin{equation}
\frac{\partial\rho}{\partial t}+\nabla\cdot\left(\rho\frac{\nabla S}{m}\right)=0.
\end{equation}

\subsection{Nelsonian mechanics: }

The above transformations are connected to Nelson's stochastic mechanics.
Nelson's approach was to assume equations {[}Eq. \ref{eq:7.5}, \ref{eq:7.6},
\ref{eq:7.1}, \ref{eq:7.2}{]} and derive the quantum mechanical
equations of motion from them by averaging, skipping the step of performing
the transformations. Averaging the two counter-propagating diffusion
equations {[}Eq. \ref{eq:7.6}{]} and {[}Eq. \ref{eq:7.2}{]} gives:
\begin{align}
\frac{\partial\rho}{\partial t}+\nabla\cdot\left(\rho\nabla\left(\frac{S'+S_{*}'}{2m}\right)\right)= & \frac{1}{2m}\nabla^{2}\rho-\frac{1}{2m}\nabla^{2}\rho\\
= & 0.
\end{align}
If we define the quantum phase $S=\frac{S'+S_{*}'}{2}$ as the average
of the forward-causal and retrocausal phases, then this equation simplifies
to the quantum continuity equation:
\begin{align}
\frac{\partial\rho}{\partial t}+\nabla\cdot\left(\rho\frac{\nabla S}{m}\right)= & 0.
\end{align}
Similarly, averaging the two Hamilton-Jacobi equations {[}Eq. \ref{eq:7.5}{]}
and {[}Eq. \ref{eq:7.1}{]} gives:
\begin{equation}
\frac{1}{2}\left(\frac{D_{*}S}{Dt}+\frac{DS}{Dt}\right)=\frac{1}{2m}\nabla S\cdot\nabla S-Q-V,
\end{equation}
which simplifies to the quantum Hamilton-Jacobi equation:
\begin{equation}
\frac{\partial S}{\partial t}+\frac{1}{2m}\nabla S\cdot\nabla S+Q+V=0.
\end{equation}
Nelson also defined the particle acceleration to be:
\begin{equation}
{\bf a}\equiv\frac{1}{2}\left(DD_{*}{\bf x}(t)+D_{*}D{\bf x}(t)\right).\label{eq:1-1-1}
\end{equation}
In our notation this is equivalent to: 
\begin{align}
{\bf a}\equiv & \frac{1}{2}\left(\frac{D\frac{1}{m}\nabla S_{*}'}{Dt}+\frac{D_{*}\frac{1}{m}\nabla S'}{Dt}\right)\\
= & -\frac{1}{m}\nabla Q-\frac{1}{m}\nabla V,
\end{align}
which is Bohm's second-order equation of motion. 

\subsection{\label{subsec:Classical-transformation-3:}Classical transformation
3: }

In addition to the transformations which give rise to Nelson's stochastic
mechanics, there are two further transformations of the classical
system which are of interest. Firstly, the classical Hamilton-Jacobi
equation: 
\begin{equation}
\frac{\partial S}{\partial t}+\frac{1}{2m}\nabla S\cdot\nabla S+V=0,
\end{equation}
and deterministic classical process: 
\begin{equation}
\frac{\partial\rho}{\partial t}+\nabla\cdot\left(\rho\frac{\nabla S}{m}\right)=0,
\end{equation}
undergo a transformation of the phase $S'=S+\frac{\hbar}{2}\log\rho$
to become a retrocausal quantum-like Hamilton-Jacobi equation: 
\begin{align}
\frac{D_{*}S'}{Dt}= & \frac{1}{2m}\nabla S'\cdot\nabla S'-Q-V,
\end{align}
with gradient of the Hamilton-Jacobi equation: 
\begin{align}
\frac{D_{*}\frac{1}{m}\nabla S'}{Dt}= & -\frac{1}{m}\nabla Q-\frac{1}{m}\nabla V,\label{eq:7.21-1}
\end{align}
and forward-causal diffusion process: 
\begin{equation}
\frac{\partial\rho}{\partial t}+\nabla\cdot\left(\rho\frac{\nabla S'}{m}\right)=\frac{1}{2m}\nabla^{2}\rho.\label{eq:7.17}
\end{equation}

\subsection{\label{subsec:Classical-transformation-4:}Classical transformation
4: }

Secondly, the classical Hamilton-Jacobi equation:
\begin{equation}
\frac{\partial S}{\partial t}+\frac{1}{2m}\nabla S\cdot\nabla S+V=0,
\end{equation}
and deterministic classical process: 
\[
\frac{\partial\rho}{\partial t}+\nabla\cdot\left(\rho\frac{\nabla S}{m}\right)=0,
\]
undergo a transformation of the phase $S_{*}'=S-\frac{\hbar}{2}\log\rho$
to become a forward-causal quantum-like Hamilton-Jacobi equation:
\begin{align}
\frac{DS_{*}'}{Dt}= & \frac{1}{2m}\nabla S_{*}'\cdot\nabla S_{*}'-Q-V,
\end{align}
with gradient: 
\begin{align}
\frac{D\frac{1}{m}\nabla S_{*}'}{Dt}= & -\frac{1}{m}\nabla Q-\frac{1}{m}\nabla V,\label{eq:7.25}
\end{align}
and retrocausal diffusion process: 
\begin{equation}
\frac{\partial\rho}{\partial t}+\nabla\cdot\left(\rho\frac{\nabla S_{*}'}{m}\right)=-\frac{1}{2m}\nabla^{2}\rho.\label{eq:7.21}
\end{equation}

\subsection{Variation of Nelson's stochastic mechanics: }

An alternative approach similar to Nelson's stochastic mechanics arises
from the two additional classical transformations described above.
Averaging the two Fokker-Planck equations {[}Eq. \ref{eq:7.17}{]}
and {[}Eq. \ref{eq:7.21}{]} gives back the quantum mechanical continuity
equation: 
\begin{equation}
\frac{\partial\rho}{\partial t}+\nabla\cdot\left(\rho\frac{\nabla S}{m}\right)=0,
\end{equation}
then averaging the two gradients of the Hamilton-Jacobi equation {[}Eq.
\ref{eq:7.21-1}{]} and {[}Eq. \ref{eq:7.25}{]} gives Bohm's equation
of motion: 

\begin{align}
{\bf a}\equiv\frac{1}{2}\left(\frac{D\frac{1}{m}\nabla S_{*}'}{Dt}+\frac{D_{*}\frac{1}{m}\nabla S'}{Dt}\right) & =-\frac{1}{m}\nabla Q-\frac{1}{m}\nabla V.\label{eq:7.28}
\end{align}
An issue however, is that adding the two Hamilton-Jacobi equations
does not give the quantum Hamilton-Jacobi equation, it gives back
the classical Hamilton-Jacobi equation: 

\begin{equation}
\frac{1}{2}\left(\frac{DS_{*}'}{Dt}+\frac{D_{*}S'}{Dt}\right)=\frac{1}{2m}\nabla S_{*}'\cdot\nabla S_{*}'+\frac{1}{2m}\nabla S'\cdot\nabla S'-Q-V,
\end{equation}
which after reversing the transformations of the phase is equal to:
\begin{equation}
\frac{\partial S}{\partial t}+\frac{1}{2m}\nabla S\cdot\nabla S+V=0.
\end{equation}

\subsection{Philosophical and practical implications }

Note that all four transformations of the classical system above give
a quantum-like Hamilton-Jacobi equation and Bohmian second-order equation
of motion in the Lagrangian reference frame of particles. Each of
the cases is intriguing in that they demonstrate a classical system
can be transformed into a representation similar to that of quantum
mechanics. However, in each case the transformed classical system
is not quite quantum mechanical. The first and second transformations
describe the stochastic Lagrangian derivative of the deterministic
phase in the Bohmian-like equation of motion, whereas in quantum mechanics
the equivalent Bohmian equations of motion describe the deterministic
Lagrangian derivative of the deterministic phase. The third and fourth
transformations meanwhile describe the stochastic Lagrangian derivative
of the stochastic phase, which is potentially more useful in terms
of physical understanding than the stochastic Lagrangian derivative
of a deterministic phase. This is the same as stochastic-transformed
version of quantum mechanics developed earlier in this paper, which
also describes the stochastic Lagrangian derivative of the stochastic
phase, and so the two theories are conceptually very similar. However
a difference is that in case of quantum mechanics the quantum potential
term $Q$ becomes $2Q$ in the transformed picture, whereas in the
transformed classical system there is only a single factor of $Q$. 

Nevertheless, this does raise the question of whether classical mechanics
is in some sense `half' of quantum mechanics. Imagine we reduced the
magnitude of the quantum potential by a factor of $\frac{1}{2}$ from
the quantum mechanical value, then the system would have an equivalent
classical representation by cases 3 and 4 of the stochastic transform
of the classical system. This fact is quite peculiar, and may have
implications for understanding quantum mechanical phenomena. 

It perhaps demonstrates that to generate entanglement, the strength
of the quantum potential needs to be greater than some critical threshold,
as taking half of this critical threshold produces only a classical
system. Why this is the case is that if the quantum potential at its
current strength is able to generate entanglement, and is then hypothetically
reduced by a factor of $\frac{1}{2}$ from this level, it would no
longer be able to generate entanglement as shown by the equivalence
to the classical representation. This indicates that the property
of being able to generate entanglement must be dependent on the strength
of the quantum potential, and hence a threshold in the strength must
have been crossed to convert the quantum behaviour to classical behaviour.
Demonstrating the existence of this threshold would be interesting,
as it could provide an alternative perspective to the usual classical
limit arguments, such as $\hbar\rightarrow0$. 

The duality between quantum and classical systems using half the quantum
potential might also have implications for computational methods of
simulating quantum systems, or performing numerical calculations.
For example, reduce the quantum potential for the target system under
investigation by a factor of half, and then simulate the reduced strength
quantum system as a classical system.

There is a possible alternative to these conclusions however; perhaps
entanglement in the non-relativistic Hamilton-Jacobi equations is
not generated through the quantum potential, but occurs because the
classical potential acting on these equations is in configuration
space, forcing the phase to take on a configuration space structure.
This would mean that although the classical and quantum mechanical
equations are structurally similar in form under the stochastic transformation,
the imposition of a configuration space potential is assumed only
to occur in the quantum mechanical system. Nevertheless, the second-order
Bohmian formulation provides insurance against this proposal. In the
Bohmian interpretation, it is not necessary to calculate the phase,
only the quantum potential. Therefore the possible configuration space
nature of the phase is not important. It is easy to apply a configuration
space potential to a single Bohmian particle configuration (and indeed
this occurs even in classical Newtonian mechanics, which is the limit
of Bohmian mechanics where the quantum potential term is zero). 

The difficult aspect of the Bohmian interpretation is the calculation
of the quantum potential. It requires taking the gradient of the probability
distribution which is formed from an ensemble of Bohmian particle
configurations. Taking the gradient of this ensemble is a non-local
procedure, and also assumes that the ensemble is physically real,
not a statistical ensemble as in classical mechanics. Once the gradient
has been calculated, it needs to be evaluated at the actual location
of the particle configuration, which again is a non-local procedure.
For these reasons, the typical Bohmian interpretation is not a free
lunch, you still have to pay for the high-dimensionality of configuration
space and the non-locality. However, if the quantum potential is reduced
by a factor of half, then there is a transformation available to an
equivalent classical system. The Bohmian interpretation of the equivalent
classical system has no quantum potential to calculate, and therefore
should avoid these problems. 

To this discussion clear, we will give an explicit example. Start
with a quantum system in the stochastic representation $S'=S+\frac{\hbar}{2}\log\rho$.
The Fokker-Planck equation {[}Eq. \ref{eq:12}{]} is:
\begin{equation}
\frac{\partial\rho}{\partial t}+\nabla\cdot\left(\rho\frac{\nabla S'}{m}\right)=\frac{1}{2m}\nabla^{2}\rho,\label{eq:7.17-1}
\end{equation}
and the corresponding Bohmian equation of motion {[}Eq. \ref{eq:5.19}{]}
is:
\begin{align}
{\bf a}= & \frac{D_{*}\frac{1}{m}\nabla S'}{Dt}=-\frac{1}{m}2\nabla Q-\frac{1}{m}\nabla V,\label{eq:7.25-1}
\end{align}
where ${\bf a}$ is the Bohmian acceleration in the Lagrangian reference
frame. Now artificially reduce the quantum potential by a factor of
2. Therefore, the Bohmian equation of motion becomes:
\begin{align}
{\bf a}= & \frac{D_{*}\frac{1}{m}\nabla S'}{Dt}=-\frac{1}{m}\nabla Q-\frac{1}{m}\nabla V.\label{eq:7.25-1-1}
\end{align}
Then perform the reverse transformation to the classical transformation
3 described in section \ref{subsec:Classical-transformation-3:}.
The Bohmian equation of motion {[}Eq. \ref{eq:7.25-1-1}{]} becomes
a classical Hamilton-Jacobi equation:
\begin{equation}
\frac{\partial S}{\partial t}+\frac{1}{2m}\nabla S\cdot\nabla S+V=0,\label{eq:7.37}
\end{equation}
while the Fokker-Planck equation {[}Eq. \ref{eq:7.17-1}{]} becomes
a deterministic continuity equation.
\begin{equation}
\frac{\partial\rho}{\partial t}+\nabla\cdot\left(\rho\frac{\nabla S}{m}\right)=0.
\end{equation}
Using $\frac{d}{dt}=\frac{\partial}{\partial t}+\frac{1}{m}\nabla S\cdot\nabla$,
the gradient of the classical Hamilton-Jacobi equation {[}Eq. \ref{eq:7.37}{]}
then becomes:
\begin{equation}
\frac{d^{2}{\bf X}(t)}{d^{2}t}=\frac{d\frac{1}{m}\nabla S}{dt}=-\frac{1}{m}\nabla V,
\end{equation}
which is a classical Newtonian second-order equation of motion for
the Bohmian particle configuration ${\bf X}(t)$. 

\section{\label{sec:Correspondence-principle-of}Subsystem correspondence
principle }

\noindent Another application of the results of this paper is to
provide a local-realist ontology for the marginal probabilities of
quantum subsystems, instead of for the total quantum system. Although
it is not widely recognised, the marginal probability distributions
of the subsystem are difficult to explain using standard quantum mechanics
\cite{key-9}, despite the subsystem for a single-particle being evidently
local.

For illustration, the subsystem idea is akin to entanglement thought
experiments involving Alice and Bob. Intuitively, Alice's subsystem
is local when considered in isolation, because any non-local information
is purely contained in the correlations to the external system which
are not observable at the level of the subsystem. A problem in the
ontology occurs however, when Alice's subsystem is highly entangled
to the external system. Although it is simple to demonstrate the subsystem
is local through the mathematical equations, as must be the case to
ensure Einsteinian local-causality, it is difficult to explicitly
construct an ontological picture of the information propagating into
the subsystem. The subsystem is like a wavefunction contained within
a closed box, only it is not a pure state but a mixed state. Entangling
interactions cause complex forms of information to flow across the
boundary of the box, which effect the mixed-state description. It
is not currently known how to describe the contents of the box and
the resulting mixed-state in a local-realist ontology. This issue
is directly connected to the more commonly studied phenomena of non-Markovinity
of quantum subsystems, which has been gaining growing attention in
recent years. 

Our investigations in this paper potentially provide a new way to
explain the ontology of quantum subsystems. Since the quantum potential
term has been removed via the stochastic transformation, only the
interactions via the classical potential need to be described for
the subsystem. This might solve the issue of having to describe a
complex suite of information flowing into the subsystem, across the
boundaries of the metaphorical box isolating a single particle. Moreover,
we note that the single particle subsystem evidently can be described
in three-dimensional space. Therefore the more difficult problem of
embedding the propagation of the configuration space wavefunction
in three-dimensional space is not present for the quantum subsystem.
The main problem remaining for the subsystem is not configuration
space or non-locality, but the presence of the quantum potential in
the mathematical description, which is solvable through the stochastic
transformation. 

This was indeed the original motivation we had for transforming the
quantum Hamilton-Jacobi equations to the Lagrangian reference frame
of stochastic particles. We hypothesized that if the quantum potential
could be removed using the stochastic transformation, the ontology
of the subsystem would simplify, and the effects of entanglement of
the subsystem to the external environment would be more natural to
describe than using the standard approach. However, it so happens
that removing the quantum potential also offers a description for
the total quantum system. Nevertheless, we should keep in mind the
interpretation of the quantum subsystems even as the ontology of the
total quantum system becomes further understood, as any issues in
the ontological description of the quantum subsystem strongly suggest
that the representation used for the total quantum system is not correct.
This line of argument is quite similar to the idea of the correspondence
principle in quantum mechanics, where is is believed that the quantum
mechanical description should reduce to the classical description
through a suitable limiting procedure, for instance taking the limit
of $\hbar\rightarrow0$. What we are suggesting is that a sensible
local-realist ontology for the quantum subsystem should be be reproducible
from the ontology of the total quantum system in some limiting procedure.
Examples of the limiting procedure might be ignoring the environmental
degrees of freedom or taking their expectation value.

Upholding the subsystem correspondence principle places important
constraints on the possible ontologies for the total quantum system.
For example, although it is possible in theory that non-locality solves
the problem of quantum entanglement, the non-locality approach toward
the total quantum system offers no further insights into the ontology
of the quantum subsystem. The quantum subsystem is evidently local,
and any interpretation that explains the results of the quantum subsystem
via non-locality has violated the subsystem correspondence principle.
This is primarily what is concerning about the non-locality approach.
To accept non-locality for the total system is to reject it unnecessarily
for the subsystem. There can only be one physical explanation for
the underlying phenomena, and it is not suitable to choose the most
convenient explanation in different situations, depending on whether
we are discussing the total quantum system or the subsystem. 

It is only by respecting the intellectual constraints of the problem
that what is being done can be considered science and not speculative
imagination. Furthermore, it is the constraints of the problem that
make it challenging, and the solution uniquely determined. Therefore,
upholding the subsystem correspondence principle appears to be of
evident importance in the ontology of quantum mechanics. The principle
is also very broad, and may be applicable to other areas in philosophy. 

\section{Conclusion }

\noindent This paper has investigated a curious set of transformations
of the quantum phase of the form $S'=S+\frac{\hbar}{2}\log\rho$ and
$S'=S-\frac{\hbar}{2}\log\rho$. These transformations convert the
deterministic quantum system of the continuity and quantum Hamilton-Jacobi
equations respectively, into a stochastic system of the Fokker-Planck
and stochastic Hamilton-Jacobi-Bellman equations. The properties of
the transformation are such that the Fokker-Planck equation remains
forward-causal, while the stochastic Hamilton-Jacobi-Bellman equation
becomes retrocausal, and vice-versa if using the retrocausal transformation
of the phase. 

We have endeavored to connect this property of the stochastic transformation
to the idea that the quantum particle propagates in the opposite temporal
direction to quantum phase. This suggestion is furthermore motivated
by the analogy between quantum mechanics and dynamic programming.
The quantum phase appears to be related to the dynamic programming
value function, while the de Broglie-Bohm particle appears to be related
to the computational solution of the dynamic programming problem.
Both the method of dynamic programming and the quantum mechanical
equations of motion are solved using a process of backward induction,
which would explain the occurrence of entanglement correlations and
non-locality. 

Attempting to embed the stochastic Hamilton-Jacobi-Bellman and Fokker-Planck
equations in three-dimensional space is however challenging. Even
if a retrocausal interpretation can solve the problem of non-locality,
it is ideally required to describe the physics in three-dimensional
space, otherwise the theory may be local, but not sufficiently realist.
Because an entangled wavefunction has a very high-dimensional state
space, it is difficult to envisage this dimensionality corresponds
to real physical objects. Nevertheless, we have attempted two ways
to embed the physical system in three-dimensional space, based on
the first and second-order versions of the de Broglie-Bohm interpretation.

In section \ref{sec:From particles to policies}, we further explored
concepts of dynamic programming in the context of quantum mechanics.
In particular we highlighted that the standard approach to the de
Broglie-Bohm interpretation is based upon value iteration. However,
dynamic programming problems can also be solved using the method of
policy iteration. We have attempted to use policy iteration to explain
how the quantum phase and gradient of the phase can acquire a configuration
space structure, yet be described in three-dimensional space. The
essence is that policy iteration is a stochastic procedure, therefore
it has a statistical ensemble of results. Taking the ensemble average
of these extended structures, of the value function and policy function
respectively, produce non-separable descriptions in configuration
space. This introduces a more general class of correlation than is
possible in the de Broglie-Bohm interpretation, which only has the
particle configurations as elements of the statistical ensemble. It
may be necessary therefore to move away from the language of particles
in the de Broglie-Bohm interpretation, and toward the language of
policy functions in the sense of dynamic programming, especially in
light of the stochastic dynamic programming formulation of the de
Broglie-Bohm interpretation. 

We have furthermore investigated the results of transformations of
the phase in the classical system. When applied to the stochastic
classical system, the transformations can be shown to be related to
Nelson's stochastic mechanics. Additionally, two transformations of
the deterministic classical system were presented. It is quite puzzling
that the deterministic continuity equation and classical Hamilton-Jacobi
equation transform to the same Fokker-Planck equation and stochastic
Hamilton-Jacobi equation as in quantum mechanics, with the only difference
being the quantum potential is reduced by a factor of $\frac{1}{2}.$
This peculiar fact may be a promising avenue for understanding new
aspects of quantum entanglement (for instance minimum thresholds of
the quantum potential required for entanglement generation), and perhaps
even provides new computational or analytical methods. 

To conclude, as Bell suggested \cite{key-8}, quantum non-locality
is not entirely forced upon us by Bell's theorem. It is possible that
a novel philosophical loophole, such as retrocausality, can be used
to evade the theorem. The questions are then, \textquotedbl What does
the high-dimensional nature of the entangled wavefunction represent
physically?\textquotedbl , \textquotedbl Can we describe quantum mechanics
in three-dimensional space or is configuration space inherent?\textquotedbl ,
and \textquotedbl Can the analogy to dynamic programming help us understand
the computational complexity of the wavefunction dynamics?\textquotedbl .
We believe the answer to these questions can be found in transforming
the quantum Hamilton-Jacobi equation and continuity equation to a
new picture, perhaps the stochastic picture we have provisionally
developed here. 

\section*{Statements \& declarations}

\subsection*{Statement of originality}

All work is original research and is the sole research contribution
of the author. 

\subsection*{Conflicts of interest: }

No conflicts of interest. 

\subsection*{Funding: }

No funding received. 

\subsection*{Copyright notice: }

Copyright{\small{} }{\footnotesize{}© }2024 Adam Brownstein under
the terms of arXiv.org perpetual, non-exclusive license.

\appendix

\section{Stochastic Lagrangian derivative \label{sec:Lagrangian-derivative-of}}

\subsection{Retrocausal stochastic Lagrangian derivative\label{subsec:Retrocausal-Lagrangian-derivativ}}

\noindent Suppose that a collection of particles are undergoing stochastic
motion according to the guidance equation:
\begin{equation}
d{\bf X}(t)={\bf v}dt+\sqrt{k}d{\bf W}_{t},
\end{equation}
where ${\bf v}$ is the velocity of the particle configuration, and
$d{\bf W}_{t}$ is a Wiener process. The definition of the retrocausal
stochastic Lagrangian derivative can be found using a Taylor series
expansion: 
\begin{align}
\frac{Df({\bf x},t)}{Dt}= & E\left[\frac{f({\bf x}+\delta{\bf x},t+\delta t)-f({\bf x},t)}{dt}\right]\\
= & E\left[\frac{\partial f({\bf x},t)}{\partial t}+\sum_{c,i}\frac{\partial f({\bf x},t)}{\partial x_{ci}}\frac{dx_{ci}}{dt}+\sum_{c,i}\sum_{d,j}\frac{1}{2}\frac{\partial f({\bf x},t)}{\partial x_{ci}\partial x_{dj}}\frac{dx_{ci}dx_{dj}}{dt}\right]\\
= & \frac{\partial f({\bf x},t)}{\partial t}+\sum_{c,i}\frac{\partial f({\bf x},t)}{\partial x_{ci}}\frac{dx_{ci}}{dt}+\sum_{c,i}\sum_{d,j}\frac{1}{2}k\frac{\partial f({\bf x},t)}{\partial x_{c,i}\partial x_{d,j}}\\
= & \frac{\partial f({\bf x},t)}{\partial t}+{\bf v}\cdot\nabla f({\bf x},t)+\frac{1}{2}k\nabla^{2}f({\bf x},t),\label{eq:20}
\end{align}
where $c,d\in\mathbb{N}\setminus\{0\}$ are particle labels, $i,j\in[1,2,3]$
are coordinate indices, and $k$ is positive-valued constant. Therefore
the retrocausal stochastic Lagrangian derivative is defined as:
\begin{align}
\frac{D_{*}}{Dt}\equiv & \frac{\partial}{\partial t}+{\bf v}\cdot\nabla+\frac{1}{2}k\nabla^{2}.
\end{align}

\subsection{Forward-causal stochastic Lagrangian derivative\label{subsec:Forward-causal-Lagrangian-deriva}}

The forward-causal stochastic Lagrangian derivative can be found using
the alternative Taylor series expansion:
\begin{align}
\frac{Df({\bf x},t)}{Dt}= & E\left[\frac{f({\bf x},t)-f({\bf x}-\delta{\bf x},t-\delta t)}{dt}\right]\\
= & E\left[\frac{\partial f({\bf x},t)}{\partial t}+\sum_{c,i}\frac{\partial f({\bf x},t)}{\partial x_{ci}}\frac{dx_{ci}}{dt}-\sum_{c,i}\sum_{d,j}\frac{1}{2}\frac{\partial f({\bf x},t)}{\partial x_{ci}\partial x_{dj}}\frac{dx_{ci}dx_{dj}}{dt}\right]\\
= & \frac{\partial f({\bf x},t)}{\partial t}+\sum_{c,i}\frac{\partial f({\bf x},t)}{\partial x_{ci}}\frac{dx_{ci}}{dt}-\sum_{c,i}\sum_{d,j}\frac{1}{2}k\frac{\partial f({\bf x},t)}{\partial x_{c,i}\partial x_{d,j}}\\
= & \frac{\partial f({\bf x},t)}{\partial t}+{\bf v}\cdot\nabla f({\bf x},t)-\frac{1}{2}k\nabla^{2}f({\bf x},t),\label{eq:20-1}
\end{align}
where $c,d\in\mathbb{N}\setminus\{0\}$ are particle labels, $i,j\in[1,2,3]$
are coordinate indices, and $k$ is positive-valued constant. Therefore
the forward-causal stochastic Lagrangian derivative is defined as:
\begin{align}
\frac{D}{Dt}\equiv & \frac{\partial}{\partial t}+{\bf v}\cdot\nabla-\frac{1}{2}k\nabla^{2}.
\end{align}

\section{\label{sec:Quantum-potential}Forms of the quantum potential }

\noindent The quantum potential can be written in the following form:
\begin{align}
Q= & -\frac{\hbar^{2}}{2m}\frac{\nabla^{2}\sqrt{\rho}}{\sqrt{\rho}}\\
= & -\frac{\hbar^{2}}{4m}\frac{\nabla\cdot\left(\rho^{-\frac{1}{2}}\nabla\rho\right)}{\sqrt{\rho}}\\
= & \frac{\hbar^{2}}{8m}\frac{\rho^{-\frac{3}{2}}\nabla\rho\cdot\nabla\rho}{\sqrt{\rho}}-\frac{\hbar^{2}}{4m}\frac{\nabla^{2}\rho}{\rho}\\
= & \frac{\hbar^{2}}{8m}\nabla\log\rho\cdot\nabla\log\rho-\frac{\hbar^{2}}{4m}\frac{\nabla^{2}\rho}{\rho}.\label{eq:113}
\end{align}
It can also be written as:
\begin{align}
Q= & -\frac{\hbar^{2}}{2m}\frac{\nabla^{2}\sqrt{\rho}}{\sqrt{\rho}}\\
= & -\frac{\hbar^{2}}{2m}\rho^{-\frac{1}{2}}\nabla\cdot\left(\sqrt{\rho}\frac{1}{\sqrt{\rho}}\nabla\sqrt{\rho}\right)\\
= & -\frac{\hbar^{2}}{2m}\frac{\nabla\sqrt{\rho}}{\sqrt{\rho}}\cdot\frac{\nabla\sqrt{\rho}}{\sqrt{\rho}}-\frac{\hbar^{2}}{2m}\nabla\left(\frac{1}{\sqrt{\rho}}\nabla\sqrt{\rho}\right)\\
= & -\frac{\hbar^{2}}{8m}\nabla\log\rho\cdot\nabla\log\rho-\frac{\hbar^{2}}{4m}\nabla^{2}\log\rho.\label{eq:113-1}
\end{align}
Adding the first {[}Eq. \ref{eq:113}{]} and second {[}Eq. \ref{eq:113-1}{]}
forms of the quantum potential gives a third form: 
\begin{align}
Q= & -\frac{\hbar^{2}}{8m}\nabla^{2}\log\rho-\frac{\hbar^{2}}{8m}\frac{\nabla^{2}\rho}{\rho}.\label{eq:126}
\end{align}

\end{document}